\documentclass[apj]{emulateapj}
\newcommand{\lamm}{$\lambda\lambda$}
\newcommand{\lam}{$\lambda$}
\newcommand{\kms}{km s$^{-1}\,$}
\slugcomment{The Astrophysical Journal, in press} 
\shorttitle{IC 418 FAINT EMISSION LINES}
\shortauthors{SHARPEE, BALDWIN, \& WILLIAMS}
\begin{document}

\title{Identification and Characterization of Faint Emission Lines in
the Spectrum of the Planetary Nebula IC 418}

\author{Brian Sharpee\altaffilmark{1} and Jack A. Baldwin} 
\affil{Department of Physics and Astronomy, Michigan State
University, East Lansing, MI 48824} 
\and
\author{Robert Williams}
\affil{Space Telescope Science Institute, 3700 San Martin Drive,
Baltimore, MD 21218}
 
\altaffiltext{1}{Present Address: SRI International, 333 Ravenswood
Ave, Menlo Park, CA 94025}

\begin{abstract}

We present high signal-to-noise echelle spectra of the compact high
surface brightness, low ionization planetary nebula IC 418. These
reveal 807 emission lines down to intensities less than
10$^{-5}$ that of H$\beta$ for which we determine widths and relative
intensities.  We show that line profiles are a valuable parameter for
making line identifications and in constraining the excitation
mechanism of the lines.  We present evidence that indicates that many
supposed high-level recombination lines may in fact be excited by a
process other than recombination.  We contend from the detection of
dielectronic recombination lines that their relatively low intensities
argue against their making a significant contribution to level
populations of the heavy ions in this object.  Following similar
analyses of other PNe we find that IC 418 shows a small discrepancy in
ion abundances derived from forbidden vs. recombination lines of the
heavy elements.

\end{abstract}


\section{Introduction} \label{intro}

A significant source of our knowledge of chemical abundances
throughout the universe is spectrophotometry of distant emission-line
regions. The classical approach (e.g. Osterbrock 1989) is to determine
the relative abundances of elements heavier than helium by comparing
the intensities of their strong, collisionally excited lines (which in
the optical regime are mostly forbidden lines) to the intensities of
the strong hydrogen recombination lines. Major results from such work
include measurements of the level of CNO enrichment in starburst
galaxies seen all the way back to redshifts z$\sim$6
\citep{SM96,H02,R03} and the abundance gradients in disk galaxies
\citep{Z94}. Direct observational measurements of the primordial
helium abundance also rest on measuring the trend of He/H vs. CNO/H in
external galaxies, and then extrapolating back to zero metallicity,
with some claims of accuracy in the third decimal place
\citep{P92,PK92,PPL02}.

An unsettling result that has unfolded over the past decade
\citep{B91,L95,L00,GD01a,T04} is that when the abundances of C, N, and
O are measured in \ion{H}{2} regions and planetary nebulae (PNe) using
the much weaker recombination emission lines of these elements, the
results are often significantly different from those obtained using
the stronger collisionally excited forbidden emission lines from the
same ions. The discrepancies are typically factors of 2--3, but for
some objects they are more than an order of magnitude, and the
recombination lines consistently yield higher abundances.

The recombination lines might be expected to be the more reliable
abundance indicators because recombination coefficients depend much
less strongly on electron temperature than do the collisional
excitation coefficients that set the intensities of the collisionally
excited lines. However, recombination lines are weak and other
processes can compete with recombination in exciting high level
lines. These include: Bowen-like fluorescence mechanisms
\citep{G76,F92}, continuum fluorescence from the ionizing stars
\citep{S68,G75a,G75b,G76,E02}, dielectronic recombination
\citep{GD01a,GD01b}; and charge exchange \citep{BD80}.  On the other
hand, the accuracy of the abundances determined from collisionally
excited lines could be affected by temperature or density fluctuations
\citep{P67,PST93,VC94,P04} or other factors that lead to calculating
forbidden line strengths using incorrect or incorrectly weighted
electron temperatures.  The existence of distinctly different
structural components within the PNe giving rise to either only
collisionally excited or only recombination lines has also been
proposed \citep{L00,T04}.  It is important to understand the source of
these abundance discrepancies because of the widespread use of these
lines as a general tool for measuring chemical abundances.

\begin{figure*}
\epsscale{0.6}
\plotone{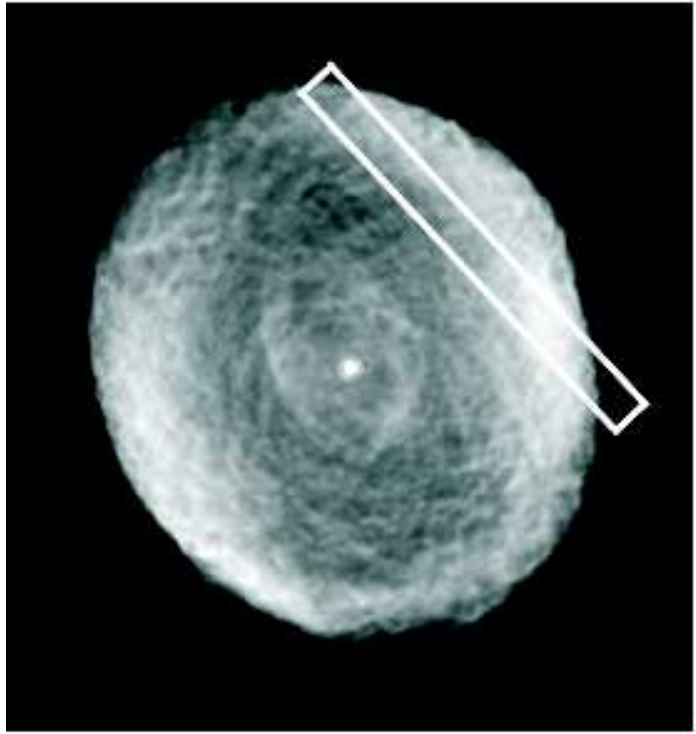}
\epsscale{1.0}
\caption{HST image of IC 418, with the position and size of the
blue/intermediate set-ups' slit marked.  The long-axis of the slit is
oriented along the north-south line with north in the direction of
the upper left-hand corner of the image and east to lower left-hand
corner.\label{f1}}
\end{figure*}

In this paper, we use deep, high-resolution spectra of the planetary
nebula IC 418, described previously in the first paper of this series
which focused on line identification (Sharpee et al. 2003; Paper I),
to investigate the behavior of hundreds of weak permitted emission
lines of C, N, O, and Ne in order to gain further insight into the
source of the abundance discrepancies. We have chosen to study IC 418
partly because of its high surface brightness, but also because of the
apparent simplicity of its projected geometry. Figure~\ref{f1}
presents an HST image of the nebula, showing its well-defined oval
shape. The original color image also clearly demonstrates that the
bright outer ring consists of gas of a much lower ionization than the
fainter inner region. Our spectra support the idea that IC 418 is to
first order a symmetrically expanding flow of gas with the faster
moving gas on the outside where the ionization is lower. This produces
a strong correlation between ionization level and line-of-sight
velocity structure, as evidenced by the emission-line profiles, which
can be used to help constrain where in the PN shell different emission
lines are formed.

It should be noted that IC 418 has a considerably lower ionization
level than most PNe due to the moderately low temperature of its
central star, so that (a) our data do not include recombination lines
from some of the higher ionization states that have been studied in
other PNe, and (b) the gas is subjected to a systematically different
(softer) continuum shape than is typical for the PNe for which the
largest abundance discrepancies have been measured. Also, since
\citet{GD01a} have shown that abundance discrepancies between
forbidden and recombination lines correlate inversely with the PN
surface brightness, and IC 418 has a relatively high surface
brightness, it may not be among the best objects to study in order to
understand this phenomenon.

\section{Observations and Basic Data Reduction}\label{obs}

We observed IC 418 for five nights using the Cassegrain echelle
spectrograph on the 4m Blanco telescope at CTIO. For the first two
nights, 2001 Dec 27-28 UT, we used a blue setup giving full wavelength
coverage over \lamm3500--5950\AA\ with the blue long camera and
optics, a 79 line mm$^{-1}$ echelle grating and a 316 line mm$^{-1}$
cross-dispersing grating blazed at 4400\AA. We then switched to the
red long camera and optics, a GG495 order separating filter, the 31.6
line mm$^{-1}$ echelle grating, and a 316 line mm$^{-1}$ cross
disperser blazed at 7500\AA. We used this configuration for two nights
with an intermediate setup that covered \lamm5090--7425\AA, followed
by a final night with a red setup covering \lamm7350--9865\AA.

For all three setups the slit width was 1\arcsec. The slit length
decker was set so as to avoid overlap of the echelle orders at the
short wavelength end of the wavelength range of each setup, and this
corresponded to 11.9\arcsec\ for the blue and intermediate setups, and
19.6\arcsec\ for the red setup.  The nebula was slightly larger than
the shorter decker length, but fit within the longer decker length.

\begin{figure*}
\plotone{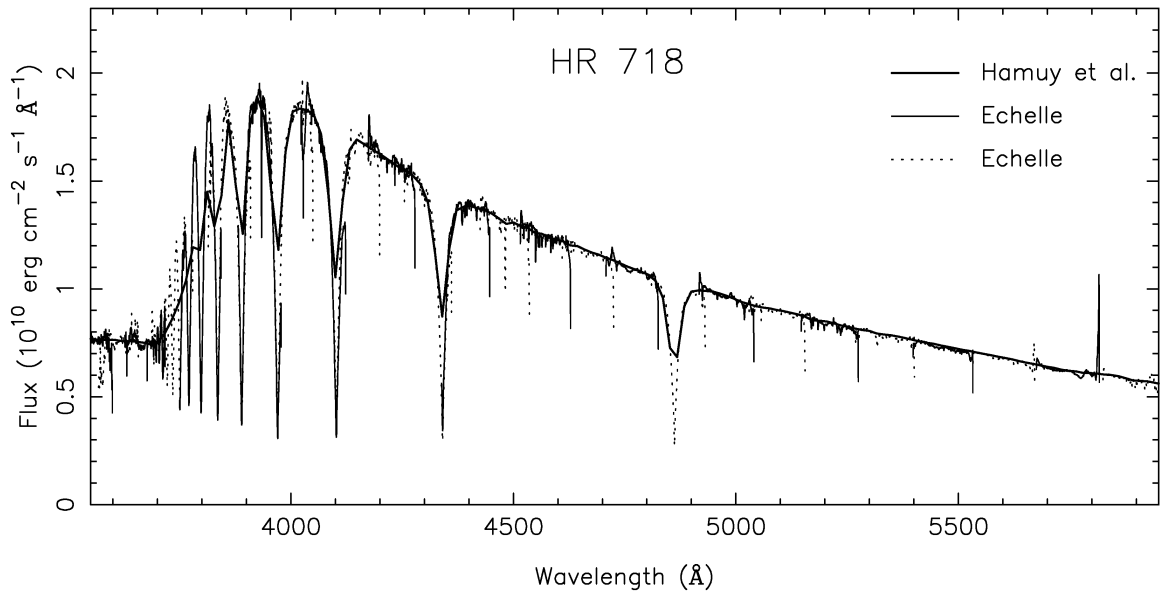}
\caption{Flux calibrated echelle
spectrum of the standard star HR 718, compared to 16\AA-resolution
calibration spectrum of Hamuy et al.\ (1994).  Successive echelle
orders are alternatively plotted with thin solid and dotted lines so
that the level of agreement between adjacent orders may be
assessed. \label{f2}}
\end{figure*}

We observed IC 418 over a span of approximately 6 hours on each
night. Most of the data were taken as a series of 1800s (and
occasionally 1000s) exposures, but shorter exposures of 30s, 120s, and
300s were also taken with each setup in order to measure the few
strongest lines which were saturated on the long exposures. The total
observing time for the co-added long exposures for each setup were:
23,600 s (blue), 30,600 s (intermediate), and 18,000 s (red) The
position of the slit on the nebula is shown in Figure~\ref{f1} for the
blue and intermediate setups. We used an autoguider to set and
maintain the slit position on the nebula, returning each night to the
same guide star and using the same offset from the central star.  Our
experience was that this method gave positioning that was reproducible
to an accuracy of less than 1\arcsec. The slit was oriented N-S in all
cases, which was the direction of the atmospheric dispersion when IC
418 crossed the meridian (at 18 degrees zenith distance) about halfway
through each night's observations. However, this was significantly
different from the direction towards the zenith at the start or end of
the observations (when IC 418 was about 3$^h$ E and 3$^h$ W of the
meridian, respectively).  The autoguider has a red sensitive camera,
so while in the red the slit position was maintained fairly well, in
the blue there was an extra smearing over about 1.5 \arcsec\ across
the face of the nebula. The short exposures were generally taken as IC
418 crossed the meridian.

The instrumental resolution was measured to be 9 \kms FWHM, which is
somewhat narrower than the 15 \kms FWHM profiles of the narrowest IC
418 emission lines with high quality profiles. We took Th-Ar
comparison spectra about once per hour. The wavelength fits to these
comparison spectra each had an RMS internal scatter of about 1 \kms,
and a mean error that was much smaller. The differential instrumental
flexure was small for the intermediate and red setups, but a loose
component in the blue optics train led to drifts of many pixels during
the nights of the blue setup. We paid careful attention to this during
the data reduction process, reducing each exposure of IC 418 to the
wavelength scale from the comparison lamp exposure taken closest to it
in time. We then made additional small shifts to bring into alignment
the mean of the measured wavelengths of many different emission lines
from each individual spectrum before co-adding all spectra taken with
a particular setup. Small velocity shifts were then made to the
co-added spectra to bring the measured wavelengths of night sky lines
to their laboratory values.

We then compared the measured wavelengths of nebular emission lines
from the overlapping portions of the different wavelength
setups. There was no shift between the intermediate and red setups,
but the blue setup had to have 1.2 \kms added to its wavelength scale
to bring it into agreement with the intermediate setup. After applying
these corrections, we believe that the measured wavelengths of
individual lines are accurate to about 1 \kms, and that the
mean wavelength calibration is significantly better than that. From
our measurements of 47 Balmer and Paschen lines, we find a
heliocentric velocity for IC 418 of +61.3 \kms. This agrees well with
the published values of +61.0 \kms \citep{A92} and +62.0 \kms
\citep{W53}.

\begin{figure*}
\plotone{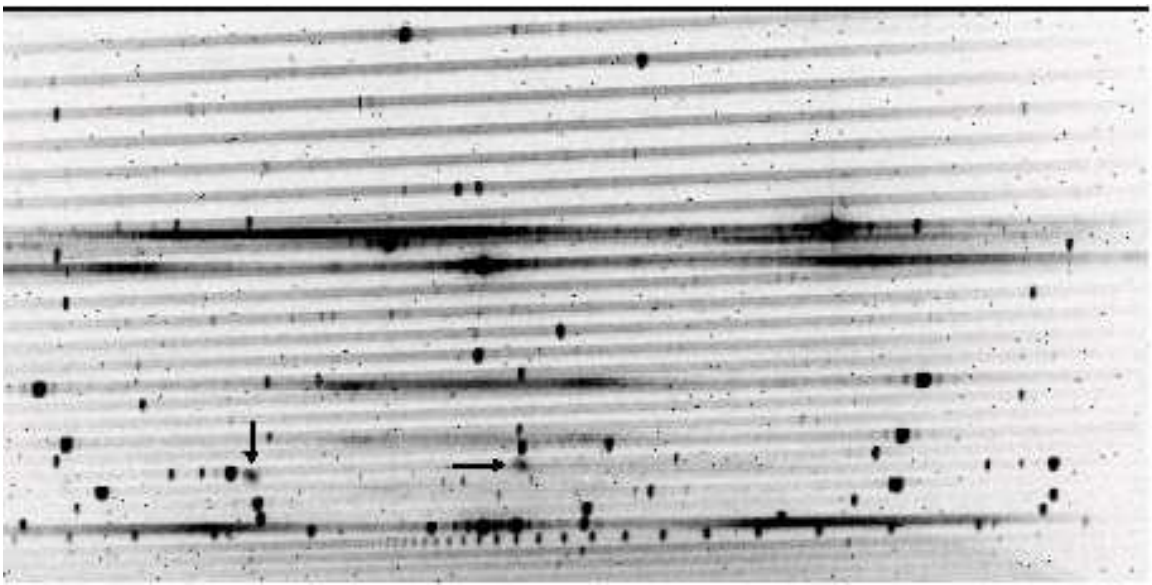}
\caption{2-D image of blue set-up IC 418 spectrum showing bright
horizontal flares that are due to scattered light from the strongest
emission lines, as well as two prominent circular ghost images
indicated by arrows.  \label{f3}}
\end{figure*}

All five nights were photometric. Flux calibrations were obtained by
observing on each night, through a 6.6\arcsec\ wide slit oriented at
the parallactic angle, three standard stars (HR718, HR4468 and HR9087)
that were calibrated at contiguous 16\AA\ intervals by
\citet{H94}. These are all late-B to early-A type white dwarfs with
very broad Balmer absorption lines, each of which fills a substantial
fraction of an echelle order. This complicates the flux calibration
because the shape of the absorption lines is folded into an
instrumental response function that has fairly sharp changes with
wavelength. We worked around this problem by also taking spectra of a
quartz continuum lamp which we divided into both the IC 418 and
standard star spectra to take out the high-order wiggles due to the
instrumental response. This permitted us to fit the flux calibration
curve using very low-order polynomials. The response curves from the
different standard stars all agreed to within a few
percent. Figure~\ref{f2} shows the combined spectra of all the orders
for one sample standard star which was reduced using the average
response curve, and it agrees well with the low-resolution spectrum
from \citet{H94}.  The overlapping parts of adjacent orders agree to
between 1--5\% except at the very edges of the orders.

In spite of the good flux calibration achieved with the standard
stars, at the much fainter light levels in the IC 418 continuum
spectrum we encountered a number of serious problems with scattered
light from the strong emission lines. We used the IRAF task
\textit{apscatter} to remove the smoothly varying scattered light
component that permeated the two-dimensional images. This worked
reasonably well except at the short-wavelength ends of each order,
where the orders were crowded too closely together to permit
measurement of the scattered light between the orders. The residuals
from subtracting off this component got multiplied by the flux
calibration curve, producing a false continuum component that curves
sharply upwards at either end of each order. In addition, there were
strong flares cutting across the orders in the vicinity of the
brightest emission lines, and various ghost images whose random nature
prevented all attempts at removal.  We could only keep track of the
positions of these artifacts and take their presence into account when
we fit the continuum shape. All of the problems related to scattered
light affected the continuum level in an additive way. This meant that
after subtracting off the continuum the emission lines intensities
could still be measured reliably to the same accuracy as the flux
calibrated standard stars except in a few regions where the scattered
light dominated the true spectrum.  We display one of the long
exposure spectra images in Figure~\ref{f3}, showing the prominence of
scattered light features in regions of the spectrum.

A number of the emission lines measured in the IC 418 spectra were
repeat measurements of the same lines occupying overlapping wavelength
ranges in adjacent echelle orders or in regions of overlap between the
different setups. The ratios of the line fluxes for these repeat
measurements afforded an additional check on the accuracy of the flux
measurements. For the ratios of lines measured in different orders
within each setup, the means of the ratios are bound to be close to
unity because the individual ratios were calculated by taking the line
flux from the redder order and dividing it by the line flux from the
bluer order. However, the scatter of the ratios about this mean gives
an idea of how reproducible the fluxes are from order to order. For
lines with S/N$>$20, the 1$\sigma$ scatter is about 20\% for the blue
setup and 9\% for the red setup. These are in fact worst-case measures
of the overall flux accuracy because they use lines that are near the
edges of the individual orders, where the flux calibration curve is
poorly determined. We also compared the flux ratios of lines measured
in more than one setup. The intermediate and red line fluxes agree
well, with a mean ratio of 0.97$\pm$0.05, whereas the mean
intermediate/blue ratio was 1.1$\pm$0.2.  We re-scaled the blue fluxes
to bring their mean into agreement with the other setups.  Another
potential check of the flux scale was to compare our results to other
published work. However, even the intensity ratios of strong lines
close together in the spectrum were very different between our work
and the data of \citet{HAF94} and \citet{HKB00}. For example, we find
the [\ion{O}{3}] \lam5007/H$\beta$ ratio to be 2.2, whereas
\citet{HAF94} measured 0.9, and \citet{HKB00} found 1.3.  These
differences are far outside the observational errors and must be due
to the three observations having been made at quite different places
in the PN shell.

A final problem was that with the blue setup the echelle grating
produced Rowland ghosts that appeared as two pairs of weak emission
lines on either side of strong, real emission lines. These were
removed following the procedure developed by \citet{B00}.

\section{Emission Line Identification}

\begin{figure*}
\plotone{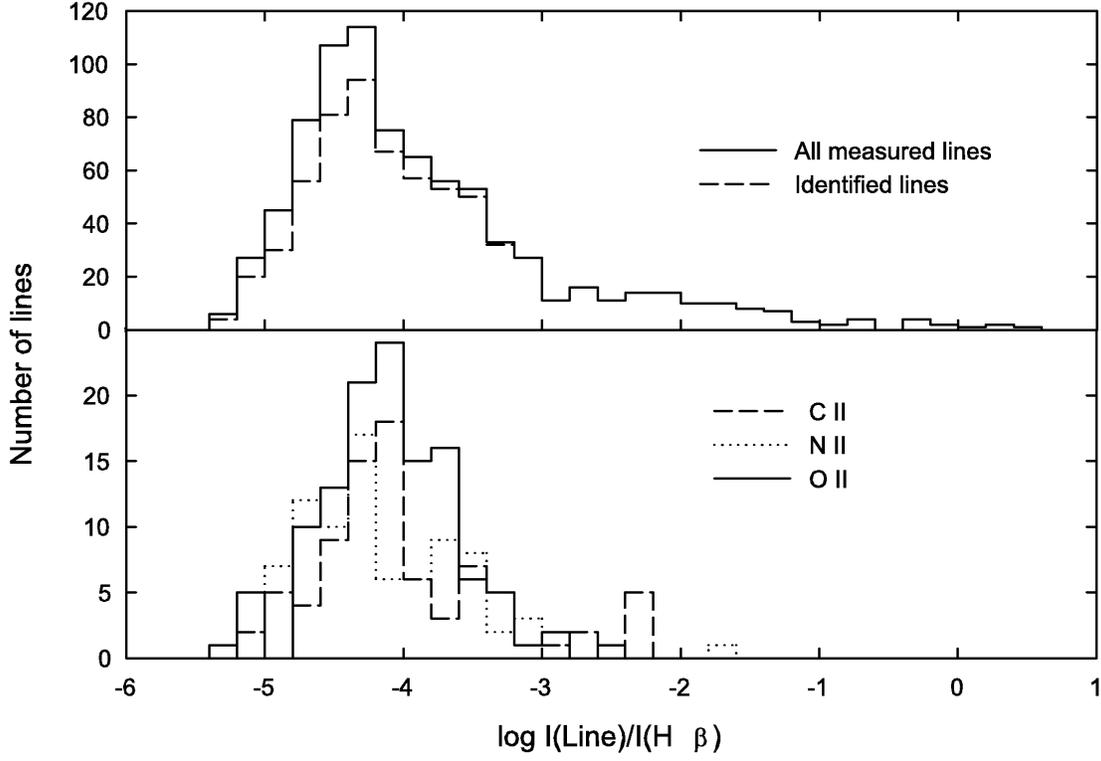}
\caption{Histograms showing the numbers of measured lines for
(upper) all lines, and all identified lines, and (lower) the permitted
lines of \ion{C}{2}, \ion{N}{2}, and \ion{O}{2}.  The numbers of lines
in the lower panel include lines for which alternate identifications
are also given, so some lines are counted more than once. \label{f4}}
\end{figure*}

\begin{deluxetable}{lclclclc}
\tablecaption{Number of line identifications per ion. \label{tab1}}
\tablehead{
 & \multicolumn{1}{c}{No.} & & \multicolumn{1}{c}{No.} & & \multicolumn{1}{c}{No.} & & \multicolumn{1}{c}{No.} \\
\multicolumn{1}{c}{Ion} & \multicolumn{1}{c}{Lines} & \multicolumn{1}{c}{Ion} & \multicolumn{1}{c}{Lines} & \multicolumn{1}{c}{Ion} & \multicolumn{1}{c}{Lines} & \multicolumn{1}{c}{Ion} & \multicolumn{1}{c}{Lines}
}
\startdata
$[$Ar III$]$ & 3  & $[$Fe III$]$ & 14 & N III       &  4 & $[$O III$]$ &  4 \\
$[$Ar IV$]$  & 2  & Fe III       &  1 & Ne I        &  4 & $[$P II$]$  &  1 \\
$[$C I$]$    & 3  & H I          & 76 & Ne II       &  8 & $[$S II$]$  &  4 \\
C I          & 1  & He I         &149 & $[$Ne III$]$&  2 & S II        & 21 \\
C II         & 83 & Mg I$]$      &  1 & $[$Ni II$]$ &  1 & $[$S III$]$ &  5 \\
$[$Cl II$]$  & 3  & Mg II        & 11 & $[$Ni III$]$&  1 & S III       &  2 \\
$[$Cl III$]$ & 6  & $[$N I$]$    &  2 & $[$O I$]$   &  3 & Si II       & 11 \\
Fe I         & 2  & N I          & 23 & O I         & 61 & Si III      &  1 \\
$[$Fe II$]$  & 24 & $[$N II$]$   &  4 & $[$O II$]$  &  6 &             &    \\
Fe II        & 2  & N II         & 85 & O II        &120 &             & 
\enddata
\end{deluxetable}

\begin{figure*}
\plotone{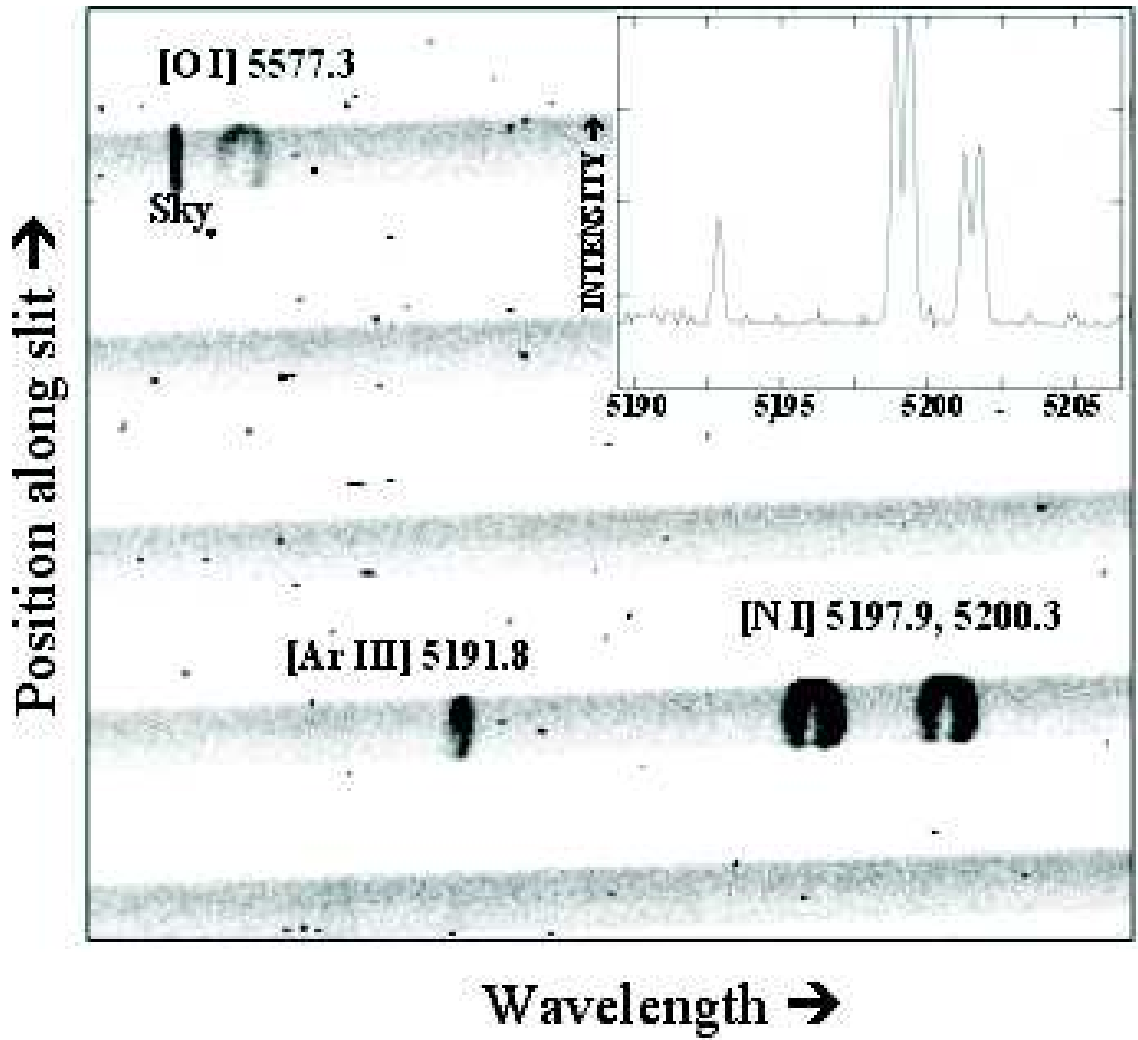}
\caption{Main figure: A portion of our
2D echelle spectrogram of IC 418, showing the characteristic profiles
of low-ionization and high-ionization lines from the PN, and of a
night sky line. Inset: Extracted spectrum showing [\ion{Ar}{3}] \lam5192
and [\ion{N}{1}] \lamm5198, 5200. \label{f5}}
\end{figure*}

\begin{figure*}
\plotone{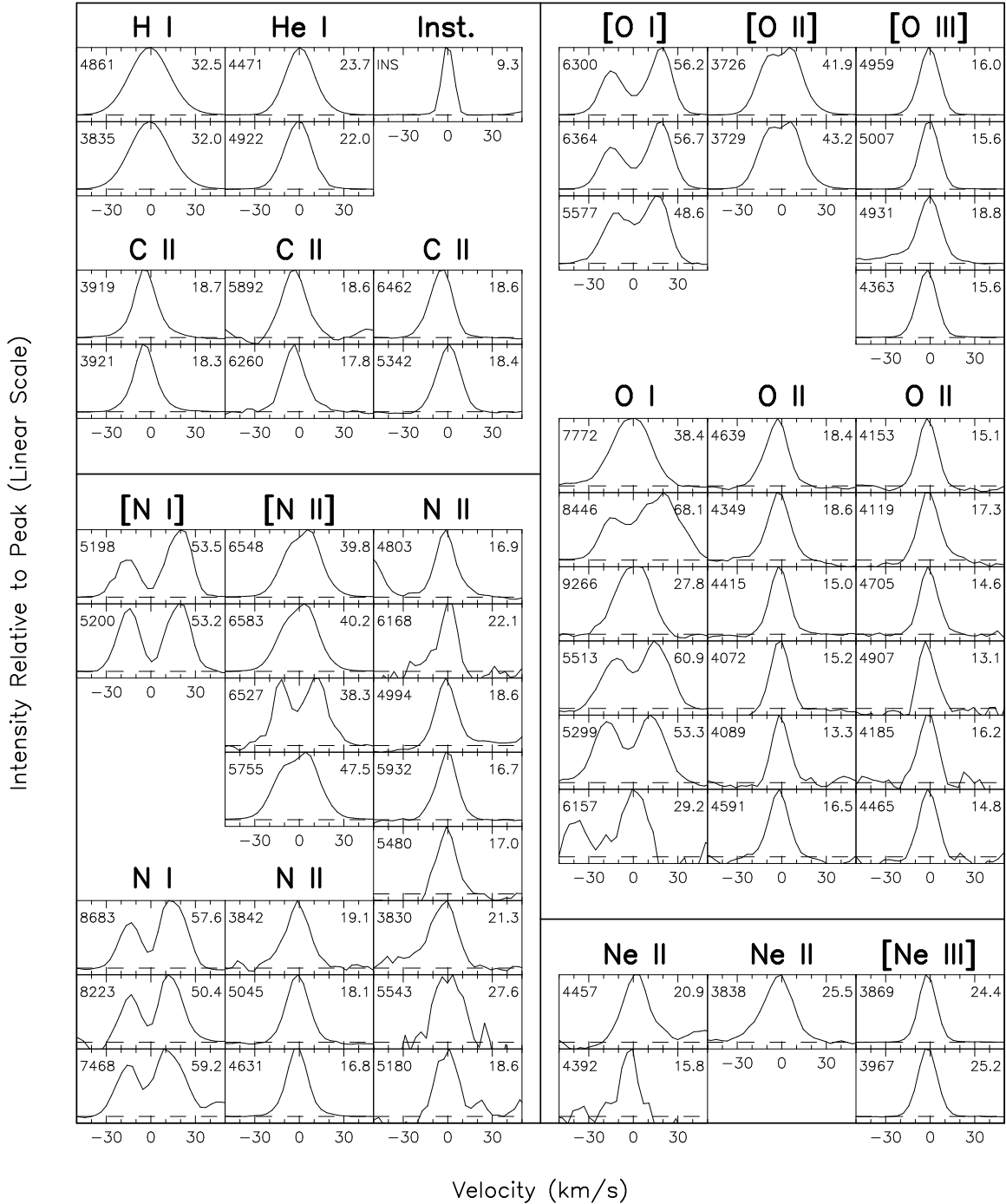}
\caption{Line profiles of unblended
lines of H, He, and C, and NONe ions for which both forbidden and
permitted lines are observed. \label{f6}}
\end{figure*}

We extracted the spectra with a specialized program that uses the
number of counts in each pixel in the original two-dimensional image
to calculate the variance in each pixel of the extracted spectrum. We
then kept track of the variance through the further reduction steps so
that we accurately knew the signal-to-noise (S/N) at each pixel in the
final, fully calibrated spectrum.  With this information, we could
then automatically search for emission lines above a consistent S/N
threshold, using the line-finding program RDGEN \citep{C01}.  The
output of RDGEN is a list of wavelengths and S/N of features
whose intensities exceed that of the continuum by a statistically
significant amount. We used a non-interactive program to fit Gaussian
profiles to those features in this list that have single-peaked
profiles, roughly half of the total number. The remaining features
have double-peaked or blended profiles, and their intensities were
determined either by the interactive deblending option in the IRAF
task \textit{splot} or, in some cases, by simply integrating the line
flux above the continuum.

These steps produced a list of accurately measured wavelengths,
widths, and fluxes for over 1523 line detections, including repeat
detections in adjacent orders and set-ups, down to a consistent S/N
limit of 7 which was the minimum value we accepted for a clearly
defined line.  An additional 52 line detections in the range
$7>$S/N$>3$ which appeared to be real features on the original
two-dimensional images were also included.  Telluric emission lines
were screened out by inspection of the two-dimensional images for
features of near instrumental resolution and uniform slit
illumination, characteristically different then the distinct profiles
of assumed nebular lines, and by comparison of wavelengths and fluxes
to the night sky emission line atlases of \citet{O96,O97}.
Elimination of likely night sky features left a list of 807 unique
emission lines from the nebula. These were dereddened using the
\citet{CCM89} reddening curves with the standard value of R$_V$
=3.1. We used 20 pairs of hydrogen Balmer and Paschen lines arising
from the same upper levels to find a reddening coefficient
$c_{H\beta}= 0.34\pm 0.05$.  This value fits in the middle of the
range of other published values including 0.25 from \citet{H80}, 0.29
from \citet{M89}, 0.21 from \citet{HAF94}, 0.34 determined by
\citet{SD95} in their re-analysis of the \citet{HAF94} data, and 0.14
from \citet{HKB00}.  This wide scatter may be due to differential
reddening across the nebula and the differing regions imaged, and/or
to the difficulty in determining an accurate reddening value due to
differing qualities of flux calibration, and numbers and types of
lines utilized in its calculation.

The final step in the data reduction process was to properly identify
the many hundreds of emission lines. We did this using EMILI, our
publicly
available\footnote{\url{http://www.pa.msu.edu/astro/software/emili/}},
automated EMIssion Line Identifier computer code which is described in
Paper I.  For each observed line, EMILI generates a list of suggested
identifications based on agreement with the expected wavelength and
flux, and on the presence of other expected lines from the same
suggested multiplet. These are the same steps that would be carried
out in a manual identification of the emission lines, but EMILI
carries them out much more rigorously and consistently. The user then
takes the lists of suggested identifications, typically with several
rank-ordered possibilities per observed line, and selects the
identifications that make the best physical sense.

This resulted in solid identifications for 624 of the 807 observed IC
418 emission lines, and possible identifications for an additional 72
lines.  The full line list is given in Table 3 of Paper I.  Here in
Table~\ref{tab1} we summarize for each ion the numbers of
identifications that have been rated in Paper I as definite.  There
are a total of 754 such line identifications, of which 480 had one
dominant candidate identification, with a further 144 observed lines
which had two or more reasonable candidate identifications and which
we interpreted as possible blends. The dereddened strengths of these
identified lines range down to slightly below 10$^{-5}$ the strength
of H$\beta$. Figure~\ref{f4} shows the distribution of all observed
lines strengths and the strengths of the permitted lines of the most
prominent heavy ions.  Further details of the observations and
reductions are given by \citet{S03}.

\section{Line Profiles} \label{profiles}

\begin{figure*}
\plotone{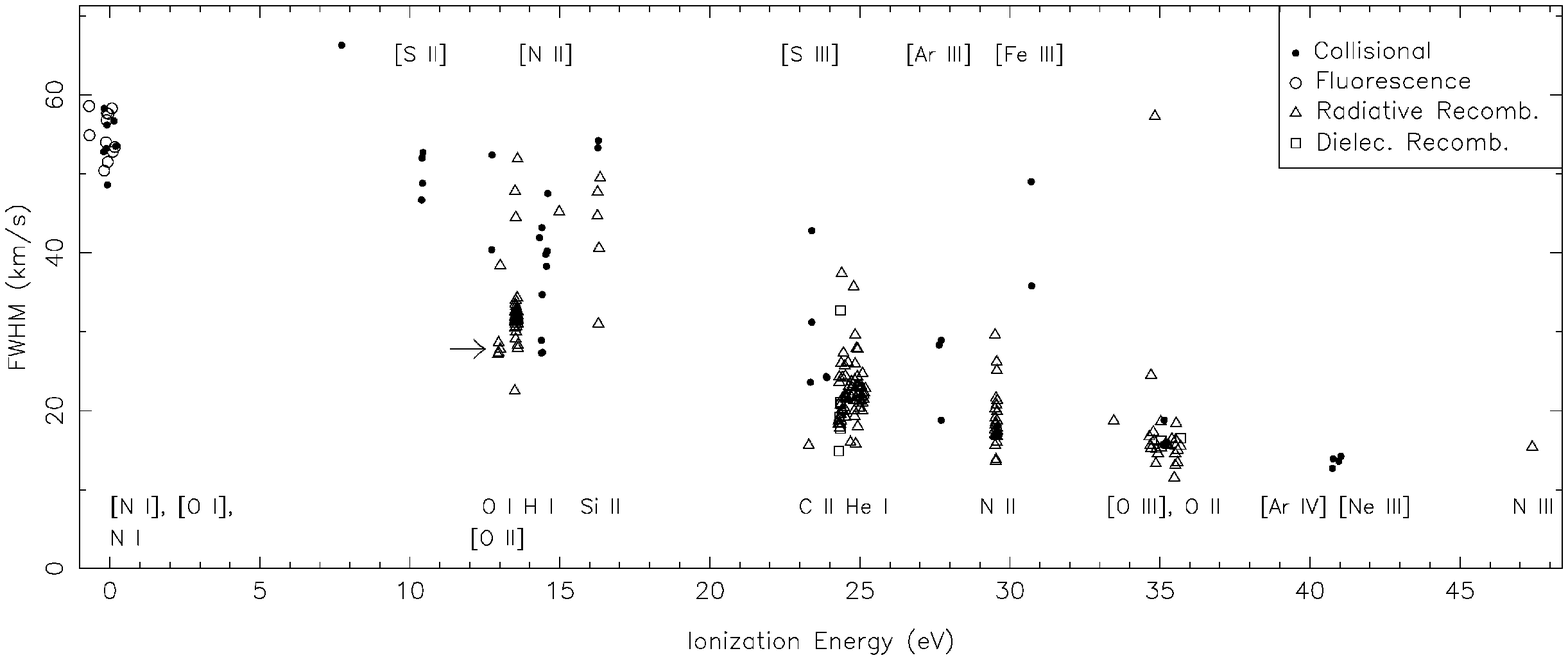}
\caption{FWHM vs. ionization energy for non-blended lines.  Closed
circles indicate collisionally excited forbidden lines, open circles
indicate lines thought to be enhanced by ground state fluorescence
processes, triangles indicate lines assumed to be due to radiative
recombination, and squares indicate lines thought to be due to
dielectronic recombination.  Radiative and dielectronic recombination
lines are plotted with the ionization energy corresponding to the
next higher ionization state than the one in which the transitions is
observed.  The arrow indicates the position of the \ion{O}{1} \lam9266
discussed in the text. \label{f7}}
\end{figure*}

As noted in \S~\ref{intro}, one of the reasons we chose to observe IC
418 at high spectral resolution is that it has a simple geometry with
a fairly large projected velocity stratification of the emission
lines. A representative segment of the two-dimensional spectrum is
shown in Figure~\ref{f5}. It is clear that the low ionization lines
from IC 418 have an oval shape, becoming strongly double peaked in the
central part of the slit, which is characteristic of the spectrum of
an expanding shell. In contrast, the higher ionization lines from the
PN are single-peaked all along the slit, which we interpret to mean
that they come from an interior part of the nebula that has a much
lower expansion velocity that was directed primarily perpendicular to
the line of sight for our slit position.
 
The discussion throughout this paper deals only with spectra extracted
over the central 6.5\arcsec\ section of the slit where the
low-ionization lines are clearly double-peaked.  Compared to spectra
that bin over the full slit length, the extracted spectra summed over
the restricted length of the slit (a) more accurately refer to the
same spot in the nebula when we compare the spectra taken with the
different setups, and (b) emphasize the profile differences between
lines coming from lower ionization regions and those coming from
higher ionization regions.  The inset in Figure~\ref{f5} shows a part
of the spectrum extracted along one of the orders that is visible in
the main figure, and illustrates the difference in line profiles
between the [\ion{N}{1}] \lamm5198, 5200 and [\ion{Ar}{3}] \lam5191
lines.  These same characteristics are seen in both the
two-dimensional images and the extracted spectra for all of the strong
forbidden lines, and are caused by the kinematical structure of the
nebula which produces different line profiles based on the location
within the nebula where the lines are formed.

In Figure~\ref{f6} we display a number of line profiles from ions with
significant numbers of permitted lines in our spectrum, and ions for
which we have measured both forbidden and permitted lines in IC 418,
where each box shows to the right of the profile the FWHM (in \kms) of
the line.  The only forbidden line profile that appears to be
inconsistent is the double-peaked structure of [\ion{N}{2}] \lam6527,
which is the weak, highly forbidden electric quadrupole transition
arising from the same upper level as the single-peaked [\ion{N}{2}]
\lamm6548, 6583 profiles.  We have no explanation for this single
discrepancy other than the possibility that this line profile may be
affected by a line blend.

The distinct profiles provide a valuable tool for diagnosing the
physical region where individual lines are formed. Consider, for
example, the profiles of permitted \ion{O}{1} \lam7772, \lam8446 and
\lam9266 from Figure~\ref{f6}.  The \lam7772 and \lam9266 lines have
profiles more similar to those of the forbidden [\ion{O}{2}] than to
the [\ion{O}{1}] lines, indicating that they are likely to be formed
in the same location as the collisionally-excited lines of the next
higher ionization state, i.e., O$^+$, as would be appropriate if their
formation were due to electron recapture. By contrast, the permitted
\lam8446 line has the same profile as the collisionally excited
[\ion{O}{1}] lines, indicating that it is formed in the O$^o$
region. In fact, \lam8446 has been shown to be predominately excited
in nebulae by resonance fluorescence from the O$^o$ ground state
\citep{G75a}.  Therefore, the profiles of individual permitted lines
give valuable information about their parent ions and the excitation
processes that produce them.  We will use this information to
constrain the excitation mechanisms for the various lines used in the
abundance analysis that follows.

\subsection{Continuum Fluorescence vs. Recombination Excitation}

Those ions for which we have definite observations of both
collisionally excited and numerous radiative recombination lines for
which cross sections are available are N$^+$, O$^+$, and O$^{+2}$.  We
use both the profiles and the strengths of the permitted lines to try
to isolate a subset of transitions that should yield reliable ion
abundances.  As illustrated in the preceding discussion, for a number
of lines that are generally agreed to be significantly enhanced by
fluorescence scattering in other PNe both the line strengths and the
line profiles indicate that this is also occurring in IC 418.
Specifically, for permitted \ion{N}{1} and \ion{O}{1} lines having the
same multiplicity as the ground state, the profiles clearly resemble
those of collisionally excited lines of the same ion, as opposed to
those of the next higher stage of ionization.
 
There are a number of other cases where Bowen-like or continuum
fluorescence are thought to contribute substantially to a line's
intensity, and this situation should be revealed by the line profile.
In order to quantify this relationship we selected lines from Table 3
of Paper I with S/N $>$ 20, and either a single ID with an
EMILI-provided Identification Index (IDI) values of five or less, or a
blend of lines from the same multiplet spaced within the instrumental
resolution of 9 \kms with at least one component meeting the IDI value
requirement.  In Figure~\ref{f7} these lines' FWHMs are plotted
against the ionization energies required to produce their parent ions.
This means for lines produced by recombination processes this is the
ionization energy necessary to produce the \textit{next higher}
ionization state than the ion the transition is observed in.  For
purposes of this plot we have assumed that all forbidden lines are
produced through electron collisional excitation, and all permitted
lines are produced either by radiative or dielectronic
recombination. Exceptions include those \ion{N}{1} and \ion{O}{1}
permitted lines from multiplets with the same multiplicity as their
respective ground terms, since Figure~\ref{f6} clearly shows that
representative lines of those multiplicities are obviously excited by
continuum fluorescence.  All \ion{C}{2} core excited doublets and all
quartets are assumed here to be produced by dielectronic recombination
\citep{D00}, as are \ion{N}{2} quintets and \ion{O}{2} sextets
(discussed further in \S~\ref{dielec}), as well as \ion{O}{2} lines
from multiplets V15, V16, and V36 (discussed further in \S~\ref{oii}).

\begin{figure*}
\plotone{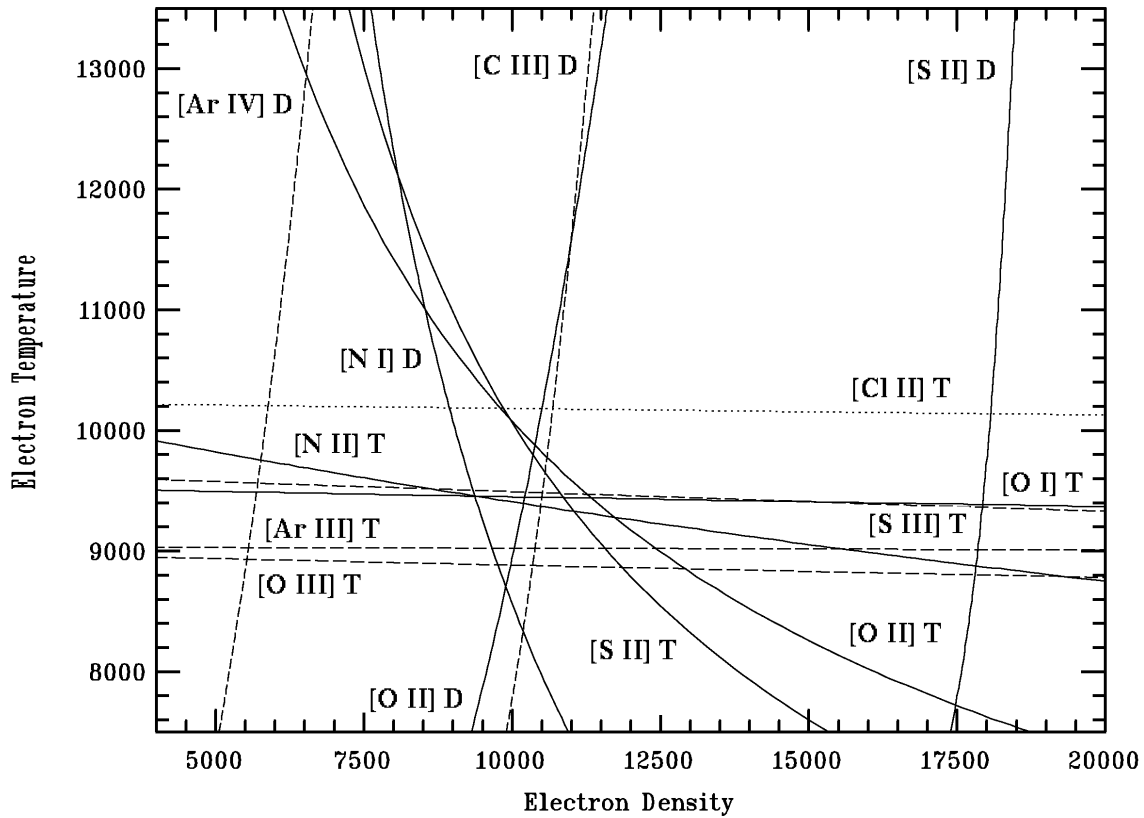}
\caption{Density-temperature diagnostic diagram for IC 418.  A ``D''
next to an ion indicates that the curve represents the density
diagnostic formed from the lines of the proceeding ion (as listed in
Table~\ref{tab2}), while a ``T'' indicates a temperature diagnostic
curve. \label{f8}}
\end{figure*}

A clear relationship is seen between line width and ionization energy,
reflecting the kinematical structure and ionization stratification of
the expanding shell.  Lines which deviate from this relation are
likely to have alternative excitation mechanisms.  For example, the
\ion{O}{1} \lam9266 line (FWHM=27.8, ionization energy=13.62 eV,
marked with an arrow in Figure~\ref{f7}), which is expected to be one
of the more prominent recombination lines, does not have the strongly
double-peaked profile of the [\ion{O}{1}] collisionally excited lines
seen in Figure~\ref{f6}, and fits well in the scheme of
Figure~\ref{f7}.  We can therefore exclude the possibility that it has
a fluorescence component, particularly since it is from a multiplet of
different multiplicity than the neutral O 2p$^4$ $^3$P ground term.
Its narrow profile indicates that it is likely formed in the interior
region of the O$^+$ ionization zone.  We note that the \ion{Si}{2}
permitted lines would actually conform better to the overall
relationship if they were assumed to be excited primarily by continuum
fluorescence (would be plotted with an ionization energy of 8.15 eV in
that case) as is shown quantitatively to be the case in the Orion
Nebula by \citet{G76}.

By using line profiles and FWHM as discriminants of the excitation
process, we can exclude those permitted lines from abundance
calculations whose strengths are significantly augmented by
fluorescence processes.

\section{Ionic Abundances in IC 418}
\subsection{Electron density and temperature}

Determining ion abundances from line strengths requires specification
of the electron temperature and density of the emitting regions.  We
have determined the electron density $N_e$ and temperature $T_e$ by
standard methods, using the intensity ratios of density-sensitive and
temperature-sensitive pairs of forbidden lines to find the locus of
possible $N_e$, $T_e$ values for the different ions represented by
each line pair, and these are shown in Figure~\ref{f8}.  We used the
IRAF package \textit{nebular} \citep{SD95} with our reddening
corrected line intensities as input, and the output consists of those
loci in the $T_e$--$N_e$ plane commensurate with each of the measured
line ratios\footnote{We used version 2.0 of the \textit{nebular}
package, but we re-installed in its database the atomic data,
spontaneous emission coefficients and collision strengths, used in the
original version which are mostly from the compilation of \citet{M83}
(but see also Table 4.1 of \citet{S03} for full details).  This is
because the revised atomic data supplied with the current version led
to a very large scatter in the $T_e$--$N_e$ values from different sets
of lines, indicating some possibly incorrect atomic data.}.

The intersection points in Figure~\ref{f8} cluster around values of
$N_e \approx 10000$ cm$^{-3}$, $T_e \approx 9600$ K, although the
[\ion{S}{2}] \lam6716/\lam6730 ratio suggests a much higher density
than any other diagnostic. This is due to the fact that the 
[\ion{S}{2}] ratio is very near its value for the high density limit,
where the ratio becomes insensitive to density, so even small
measurement errors translate into large errors in density. The
[\ion{N}{1}] \lam5200/\lam5198 intensity ratio is also near its
high-density limit, so its apparent agreement with other density
measures may be fortuitous.  The density inferred from the
[\ion{Ar}{4}] \lam4711/\lam4740 ratio is a bit discrepant from the
other values, and suggests that these relatively high ionization lines
originate in a significantly lower density region of the nebula.

\begin{deluxetable*}{lrrr}
\tablecaption{Plasma diagnostic values/uncertainties. \label{tab2}}
\tablehead{
\colhead{Lines - $\lambda$(\AA)} & \colhead{Value} & \colhead{+ Unc} & \colhead{- Unc}
}  
\startdata
\sidehead{Density $N_e$ (cm$^{-3}$)}
$[$N I$]$ $\lambda$5198/$\lambda$5200    &  9000 & \nodata & -6000 \\
$[$S II$]$ $\lambda$6716/$\lambda$6730   & 17000 & \nodata & -9000 \\
$[$O II$]$ $\lambda$3726/$\lambda$3729   & 10000 & +17000 &  -5000 \\
$[$Cl III$]$ $\lambda$5517/$\lambda$5537 & 11000 &  +4000 &  -2000 \\
$[$Ar IV$]$ $\lambda$4711/$\lambda$4740  &  6000 & +10000 &  -4000 \\
\sidehead{Temperature $T_e$ (K)}
$[$O I$]$ ($\lambda$6300 + $\lambda$6363)/$\lambda$5577                     &  9400 &  +600 & -400 \\
$[$S II$]$ ($\lambda$6716 + $\lambda$6731)/($\lambda$4068 + $\lambda$4076)  &  7000 & +4000 & \nodata \\
$[$Cl II$]$ ($\lambda$8579 + $\lambda$9124)/$\lambda$6162                   & 10200 &  +800 &  -600   \\
$[$O II$]$ ($\lambda$3726 + $\lambda$3729)/($\lambda$7320 + $\lambda$7330)  & 10000 & +4000 & -3000   \\
$[$N II$]$ ($\lambda$6548 + $\lambda$6583)/$\lambda$5755                    &  9400 &  +900 & -1400   \\
$[$S III$]$ ($\lambda$9069 + $\lambda$9532)/$\lambda$6312                   &  9500 &  +700 &  -500   \\
$[$Ar III$]$ ($\lambda$7136 + $\lambda$7751)/$\lambda$5192                  &  9000 &  +500 &  -400   \\
$[$O III$]$ ($\lambda$4959 + $\lambda$5007)/$\lambda$4363                   &  8900 &  +400 &  -400   \\
\enddata
\end{deluxetable*}

\begin{deluxetable*}{llrlrlc}
\tabletypesize{\footnotesize}
\tablecaption{Ionic abundances from collisionally excited lines \label{tab3}}
\tablehead{
\colhead{$N^{+i}$/H$^+$} & \multicolumn{1}{l}{Line(s) - $\lambda$(\AA)} & \multicolumn{2}{c}{$T_e$(K)} & \multicolumn{2}{c}{$N_e$(cm$^{-3}$)} & \multicolumn{1}{c}{Value ($N^{+i}/H^{+}$)\tablenotemark{1}}
}
\startdata
C$^o$/H$^+$ & [C I] $\lambda$8727\tablenotemark{2}                              & 9400 & [O I] & 9000 & [N I] & 1.6(-7)\\
N$^o$/H$^+$ & [N I] $\lambda\lambda$5198,5200                  & 9400 & [O I] & 9000 & [N I] & 8.6(-7)\\
N$^+$/H$^+$ & [N II] $\lambda$5755,$\lambda$6548,$\lambda$6583 & 9400 & [N II] & 10000 & [O II] & 4.1(-5)\\
O$^o$/H$^+$ & [O I] $\lambda\lambda$6300,6363                  & 9400 & [O I] & 9000 & [N I] & 5.8(-6)\\ 
O$^{+}$/H$^+$ & [O II] $\lambda$3726,$\lambda$3729,$\lambda$7320,$\lambda$7330 & 10000 & [O II] & 10000 & [O II] & 1.7(-4)\\
O$^{+2}$/H$^+$ & [O III] $\lambda$4363,$\lambda$4959,$\lambda$5007 & 8900 & [O III] & 6000 & [Ar IV] & 1.2(-4)\\
Ne$^{+2}$/H$^+$ &[Ne III] $\lambda\lambda$3869,3969             & 8900 & [O III] & 6000 & [Ar IV] & 4.3(-6)\\
S$^+$/H$^+$ & [S II] $\lambda$4068,$\lambda$4076,$\lambda$6716,$\lambda$6731 & 7000 & [S II] & 17000 & [S II] & 1.8(-6)\\
S$^{+2}$/H$^+$ & [S III] $\lambda$6312,$\lambda$9069,$\lambda$9532 & 9500 & [S III] & 11000 & [Cl III] & 2.4(-6)\\
Cl$^{+}$/H$^+$ & [Cl II] $\lambda$6162,$\lambda$8579,$\lambda$9124\tablenotemark{3} & 10200 & [Cl II] & 10000 & [O II] & 1.3(-8) \\
Cl$^{+2}$/H$^+$ & [Cl III] $\lambda\lambda$5517,5537    &  9500 & [S III] & 11000 & [Cl III] & 5.0(-8)\\
Ar$^{+2}$/H$^+$ & [Ar III] $\lambda$5192,$\lambda$7135,$\lambda$7751 & 9000 & [Ar III] & 10000 & [Cl III] & 1.0(-6) \\
Ar$^{+3}$/H$^+$ & [Ar IV] $\lambda\lambda$4711,4740 & 8900 & [O III] & 6000 & [Ar IV] & $\,$9.0(-10) \\
\enddata

\tablenotetext{1}{Value in parenthesis is the power of ten by which the
abundance should be multiplied.}
\tablenotetext{2}{Calculated using task \textit{ionic}.}
\tablenotetext{3}{Ion not implemented in \textit{nebular}.  Instead
used task \textit{ionic} for each line individually, then averaged.}

\end{deluxetable*}

We were not able to measure an accurate temperature from the Balmer
jump, because the continuum level at wavelengths longer than rest
\lam3648\AA\ was very poorly determined due to the scattered light
problems described in \S~\ref{obs}. For various attempts at fitting
the continuum shape we found Balmer jump temperatures in the range
$T_e \approx$ 5000--7000 K.

\subsection{Forbidden Lines}

The identifications and measured intensities of the collisionally
excited forbidden lines of C, N, O, Ne, S, Ar, and Cl are unambiguous,
and it is straightforward to use them to determine the corresponding
ion abundances relative to H$^+$.  We again used the IRAF
\textit{nebular} package, which includes the \textit{ionic} and
\textit{abundance} tasks for calculating abundances from the
intensities of individual lines and groups of lines,
respectively. These tasks solve multi-level atom models to find the
populations of the upper levels of emission lines for given input
values of $N_e$ and $T_e$. The calculations were made using the
density and temperature appropriate for each ionization state as
specified in Table~\ref{tab2}.  The mean values of the resulting ion
abundance ratios derived from the IRAF \textit{nebular} package tasks
using these parameters are given in Table~\ref{tab3}.

\subsection{Recombination Lines}

Due to the combination of high spectral resolution and the low
ionization of IC 418 (and hence lower density of emission lines in its
spectrum), our echelle spectra reveal numerous individual unblended
permitted emission lines from many ionic species.  To compute ionic
abundances from them it is necessary to isolate a subset of them most
likely to be excited purely by radiative recombination.  Since
permitted lines are generally weak, they could be subject to competing
mechanisms of excitation, such as fluorescence and dielectronic
recombination, as well as radiative recombination.  The relative
intensities of these lines, and the established line profile and FWHM
versus ionization energy relationships, allow us to gauge the
possibility that these two processes influence permitted line
intensities, and to choose the best subset of radiative recombination
lines for comparison in abundance with their collisionally excited
counterparts from the same ions.

\begin{deluxetable}{lll}
\tablecaption{Effective recombination coefficient references. \label{tab4}}
\tablehead{
\colhead{Lines} & \colhead{Source} & \colhead{Transitions\tablenotemark{1}}
}
\startdata
H I & Aller (1984)  & \nodata \\
He I & Smits (1996) & \nodata \\
C II & Davey et al.\ (1999) & \nodata \\
N I & P\'{e}quignot, Petitjean, \& Boisson (1991) & \nodata \\
N II & Victor \& Escalante (1990) & \nodata \\
     & Kisielius \& Storey (2002) & (3-3) \\
O I & P\'{e}quignot, Petitjean, \& Boisson (1991) & \nodata \\
O II & Storey (1994) & (3s--3p) \\
    & Liu et al.\ (1995) & (3p--3d,3d--4f) \\
    & Nussbaumer \& Storey (1984) & Mult 15,16,36 \\
Ne II & Kisielius et al.\ (1998) & (3--3) \\
      & Liu et al.\ (2000) & (3d--4f) 
\enddata 
\tablenotetext{1}{No entry in column indicate that reference was used
for all transitions.}
\end{deluxetable}

To create such a subset, Table 3 of Paper I was examined for permitted
line IDs meeting several criteria: (1) lines with a FWHM $<$ 40 \kms
for \ion{C}{2}, \ion{N}{2}, \ion{O}{2}, and \ion{Ne}{2}, the upper
limit of these ion's FWHM as seen in Figure~\ref{f7}, (2) an IDI value
less than or equal to five, or the highest ranked line among all
possible IDs suggested by EMILI, (3) no reasonable alternate IDs with
an IDI value within two of the primary ID's value, (4) multiple
detections from the same multiplet, (5) available multiplet effective
recombination coefficients.  A few additional lines were also included
where it was felt that their ID was better than the primary EMILI ID,
based upon the relative numbers of other lines in the spectrum
belonging to each ID's parent ion, and upon the levels of excitation
energy for the line behind each ID relative to those of other
well-known lines.  In cases involving likely blends between lines from
well-represented species with mutual IDI values violating condition
(3), intensities relative to the strongest line in each line's parent
multiplet were compared (by ratio of lines' effective recombination
coefficients) to determine the weaker contributor and its predicted
intensity.  Its contribution was then subtracted from the observed
intensity of the blend and the stronger component recorded in it's
parent ion's table.  The weaker component was not recorded.  If the
blend constituted the entire multiplet or if the blend was obvious (as
for \ion{O}{1} and \ion{C}{2} lines) this condition was relaxed.  This
led to a set of 16 \ion{C}{2}, 14 \ion{N}{1}, 51 \ion{N}{2}, 8
\ion{O}{1}, 57 \ion{O}{2}, and 2 \ion{Ne}{2} lines.

Abundances for these lines with respect to hydrogen were determined
utilizing the relationship,
\begin{eqnarray}
\label{eq1}
\frac{N^{+i}}{H^+} & = & \frac{\alpha_{eff}(H\beta)}{\alpha_{eff}(\lambda)}
\frac{\lambda}{\lambda(H\beta)} \frac{I(\lambda)}{I(H\beta)}\,,
\end{eqnarray}
where $\alpha_{eff}(H\beta)$ and $\alpha_{eff}(\lambda)$ are the
effective recombination coefficients for H$\beta$ and the line under
consideration, respectively.  Effective recombination coefficients for
the lines' parent multiplets were obtained from the sources listed in
Table~\ref{tab4}.  Values for the coefficients were calculated from
these references' temperature-dependent analytic fit equations using the
appropriate ion-specific temperatures listed in Table~\ref{tab5} and
assuming $N_e$=10000 cm$^{-3}$.  A ``branching ratio'', $B(\lambda)$,
a multiplicative factor equaling the ratio of an individual line's
strength to the sum of strengths of all $i$ lines in its parent
multiplet,
\begin{eqnarray}    
\label{eq2}
B(\lambda) & = & \frac{(2J+1)A\lambda^3}{\sum_i
(2J_i+1)A_i\lambda_i^3} \,,
\end{eqnarray}
where $(2J+1)$ are upper level statistical weights and $A$ are
spontaneous emission coefficients, was calculated and applied to the
multiplet's effective recombination coefficient to yield individual
line values $\alpha_{eff}(\lambda)$.  Equation 6 of \citet{K98}
computes identical branching ratios as the method above and was used
where $A$ values were unavailable.  Both methods assume that fine
structure levels are populated according to their statistical weights.
\citet{L95}, \citet{KS02}, and \citet{K98}, respectively, have noted
that $LS$-coupling inadequately describes the \ion{O}{2}, \ion{N}{2},
and \ion{Ne}{2} 3d--4f configuration transitions.  For \ion{O}{2},
individual line effective recombination coefficients computed by
\citet{L95} under an intermediate coupling case were used for the
3p--3d, and 3d--4f transitions.  Similar calculations are not
available for either of the other ions.  Therefore, the \ion{N}{2}
3d--4f lines were assigned to $LS$-coupled multiplets, and multiplet
effective recombination coefficients were calculated assuming the
LS-coupling based values from \citet{EV90}.  The \ion{Ne}{2} 3d--4f
line effective recombination coefficients were taken from the results
of \citet{L00} for NGC 6153, who used unpublished sources, so we have
accepted the small error introduced by the slight difference in
electron temperatures utilized for the two PNe.

\begin{deluxetable}{lrl}
\tablewidth{2.5in}
\tablecaption{Temperatures used for
recombination line abundance calculations\tablenotemark{1}. \label{tab5}}
\tablehead{
\colhead{Ion} & \multicolumn{1}{c}{$T_e$ (K)} & \multicolumn{1}{c}{Source} 
}
\startdata
C II & 9500  & [S III] \\
N I  & 9400  & [N II] \\
N II & 9000  & [Ar III] \\
O I  & 10000 & [O II] \\
O II & 8900  & [O III] \\
Ne II & 8900 & [O III] \\
\enddata
\tablenotetext{1}{Used a $N_e$ of 10000 cm$^{-3}$ for all cases.}
\end{deluxetable}

The abundances determined from this set of lines are recorded in
Tables~\ref{tab6}--\ref{tab9}.  Multiplet numbers in column 2 of each
table were taken from \citet{HH95}, with remaining information coming
from Table 3 of Paper I or calculated here.  Line intensities given in
italics indicate that a correction was made for blending of the type
described above.  To compute the abundances labeled as ``Sum'' for
each multiplet or configuration, the summed intensities and sum of the
ratios of line effective recombination coefficient to tabulated
wavelength for all observed multiplet lines were used in
Equation~\ref{eq1}.  Abundance values were calculated under three
different assumptions about line optical depths where relevant
effective recombination coefficients were available in the literature,
viz., Case A (all transitions optically thin), Case B (resonance
transitions optically thick), and Case C (transitions terminating on
the ground configuration O$^+$ 2p$^2$ $^2$D$^o$ term also optically
thick).  In general, Case B applies to lines from levels having the
same spin as the ground state, and Case A applies to lines of other
multiplicities.

\subsection{Notes on Individual Lines}

\begin{deluxetable*}{lccrrcrrc}
\tablecolumns{9}
\tablewidth{6.5in}
\tabletypesize{\footnotesize}
\tablecaption{C$^{+2}$/H$^+$ Abundances From Recombination Lines. \label{tab6}}
\tablehead{
 & & & & & & \multicolumn{2}{c}{C$^{+2}$/H$^+$} & \\
\multicolumn{1}{c}{Line(s)} & & & & \multicolumn{1}{c}{FWHM} & I(\lam)/I(H$\beta$) & \multicolumn{2}{c}{($10^{-4}$)} & \\ \cline{7-8}
\multicolumn{1}{c}{(\AA)} & \colhead{Mult} & \colhead{IDI/Rank} & \multicolumn{1}{c}{S/N} & \multicolumn{1}{c}{(km/s)} & (I(H$\beta$)=100) & \multicolumn{1}{c}{Case A} & \multicolumn{1}{c}{Case B} & Notes 
}
\startdata
\multicolumn{9}{l}{\textbf{V2 (3s $^2$S--3p $^2$P$^o$)}} \\
6578.050       & V2    & 2 A & 870.5 & 18.3 & 0.5374 &  56.5 &  10.3 & \nodata \\
\multicolumn{9}{l}{\textbf{V3 (3p $^2$P$^o$--3d $^2$D)}} \\
7231.340       & V3    & 3 A & 255.3 & 23.6 & 0.1692 & 294.4 &   4.2 & \nodata \\
7236.420       & V3    & 2 A & \nodata& 21.1 & 0.4673 & 451.5 &   6.4 & \nodata \\
7237.170       & V3    & \nodata &  15.5 & 16.6 & 0.0489 & 425.3 &   6.0 & \nodata \\
Sum:           & V3 & \nodata & \nodata & \nodata & 0.6854 & 397.4 &   5.6 & \nodata \\
\multicolumn{9}{l}{\textbf{V4 (3s $^2$P$^o$--4s $^2$S)}} \\
3918.967       & V4    & 4 A & 226.4 & 18.7 & 0.1068 & 182.3 &  59.4 & \nodata \\
3920.682       & V4    & 3 A & 372.8 & 18.3 & 0.2052 & 175.3 &  57.1 & \nodata \\
Sum:           & V4 & \nodata & \nodata & \nodata & 0.3120 & 177.6 &  57.9 & \nodata \\
\multicolumn{9}{l}{\textbf{V5 (3d $^2$D--4p $^2$P$^o$)}} \\
5891.600       & V5    & 5 A &  99.2 & 18.6 & 0.0181 &  33.0 &  14.1 & \nodata \\
\multicolumn{9}{l}{\textbf{V6 (3d $^2$D--4f $^2$F$^o$)}} \\
4267.001,183,261 & V6  & 5 A & 536.0 & 38.7 & 0.5712 &   5.6 &   5.5 & \nodata \\
\multicolumn{9}{l}{\textbf{V10.03 (4p $^2$P$^o$--5d $^2$D)}} \\
6257.180       & V10.03& 4 A & \nodata& 19.9 & 0.0052 & 172.6 &  12.5 & \nodata \\
6259.560       & V10.03& 4 A &  63.7 & 17.8 & 0.0084 & 155.0 &  11.3 & \nodata \\
Sum:           & V10.03 & \nodata & \nodata & \nodata & 0.0136 & 161.3 &  11.7 & \nodata \\
\multicolumn{9}{l}{\textbf{V12.01 (4p $^2$P$^o$--6d $^2$D)}} \\
4637.630       & V12.01& 8 C &  19.5 & 23.2 & 0.0056 & 104.5 & 9.1 & \nodata \\
\multicolumn{9}{l}{\textbf{V16.04 (4d $^2$D--6f $^2$F$^o$)}} \\
6151.270,0.540 & V16.04& 3 A & 127.3 & 26.7 & 0.0253 &   5.9 &   5.8 & \nodata \\
\multicolumn{9}{l}{\textbf{4d $^2$D--8f$^2$F$^o$}} \\
4620.185       &\nodata& 7 A &  41.7 & 21.5 & 0.0107 &   5.9 &   5.8 & \nodata \\ 
\multicolumn{9}{l}{\textbf{4d $^2$D--9f$^2$F$^o$}} \\
4329.675       &\nodata& 6 B &  20.7 & 21.4 & 0.0070 &   5.4 &   5.3 & \nodata \\ 
\multicolumn{9}{l}{\textbf{V17.04 (4f $^2$F$^o$--6g$^2$G)}} \\
6461.950       & V17.04& 6 C &  93.5 & 18.6 & 0.0584 &   5.5 & \nodata & \nodata \\
\multicolumn{9}{l}{\textbf{V17.06 (4f $^2$F$^o$--7g$^2$G)}} \\
5342.370       & V17.06& 4 A & 153.2 & 18.4 & 0.0277 &   5.1 & \nodata & \nodata \\
\multicolumn{9}{l}{\textbf{V17.08 (4f $^2$F$^o$--8g$^2$G)}} \\
4802.740       & V17.08& 6 B &  72.8 & 19.5 & 0.0183 &   5.7 & \nodata & \nodata \\
\multicolumn{9}{l}{\textbf{4f $^2$F$^o$--9g$^2$G}} \\
4491.130       &\nodata& 5 B &  56.5 & 19.6 & \textit{0.0109} &   5.2 & \nodata & \tablenotemark{1} \\
\multicolumn{9}{l}{\textbf{4f $^2$F$^o$--10g$^2$G}} \\
4292.250       &\nodata& 6 B &  35.2 & 21.2 & \textit{0.0079} &   5.5 & \nodata & \nodata \\ 
\enddata

\tablenotetext{1}{\ion{O}{2} 3d $^2$P$_{3/2}$--4f D $^2$[3]$^o_{5/2}$
\lam4491.222 (V86a) had smaller IDI=4 but contributes only slightly to
\ion{C}{2} 4f $^2$F$^o$--9g $^2$G \lam4292.250 with IDI=5.}




\end{deluxetable*}

\begin{deluxetable*}{lccrrcrr@{\extracolsep{0.2pc}}r@{\extracolsep{0pc}}rc}
\tablecolumns{11}
\tablewidth{6.9in}
\tabletypesize{\footnotesize}
\tablecaption{N$^{+2}$/H$^+$ Abundances From Recombination Lines. \label{tab7}}
\tablehead{
 & & & & & & \multicolumn{4}{c}{N$^{+2}$/H$^+$} & \\
 & & & & & & \multicolumn{4}{c}{($10^{-4}$)} & \\
\multicolumn{1}{c}{Line(s)} & & & & \multicolumn{1}{c}{FWHM} & I(\lam)/I(H$\beta$) & \multicolumn{2}{c}{EV90\tablenotemark{1}} & \multicolumn{2}{c}{KS02\tablenotemark{2}} & \\ \cline{7-8}\cline{9-10}
\multicolumn{1}{c}{(\AA)} & \colhead{Mult} & \colhead{IDI/Rank} & \multicolumn{1}{c}{S/N} & \multicolumn{1}{c}{(km/s)} & (I(H$\beta$)=100) & \multicolumn{1}{c}{Case A} &
\multicolumn{1}{c}{Case B} & \multicolumn{1}{c}{Case A} & \multicolumn{1}{c}{Case B} & Notes 
}
\startdata
\multicolumn{11}{l}{\textbf{V3 (3s $^3$P$^o$--3p $^3$D)}} \\*
5666.629       & V3   & 1 A & 176.9 & 17.1 & 0.0414 &   3.8 &   3.2 &   2.4 &   1.9 & \nodata \\
5676.017       & V3   & 1 A & 123.4 & 17.6 & 0.0197 &   4.1 &   3.5 &   2.5 &   2.1 & \nodata \\
5679.558       & V3   & 2 A & 170.8 & 20.7 & 0.0674 &   3.3 &   2.8 &   2.1 &   1.7 & \nodata \\
5686.212       & V3   & 2 A &  70.1 & 18.5 & 0.0127 &   3.5 &   3.0 &   2.2 &   1.8 & \nodata \\
5710.766       & V3   & 1 A &  75.9 & 18.7 & 0.0136 &   3.8 &   3.2 &   2.4 &   1.9 & \nodata \\
5730.656       & V3   & 2 A &   8.9 & 17.5 & 0.0013 &   5.4 &   4.6 &   3.4 &   2.9 & \nodata \\*
Sum:           & V3   & \nodata & \nodata & \nodata & 0.1561 &   3.6 &   3.1 &   2.2 &   1.8 & \nodata \\
\multicolumn{11}{l}{\textbf{V4 (3s $^3$P$^o$--3p $^3$S)}} \\*
5045.099       & V4   & 3 B & 112.9 & 18.1 & 0.0289 &  68.0 &  11.0 &  33.6 &   4.8 & \tablenotemark{3} \\    
\multicolumn{11}{l}{\textbf{V5 (3s $^3$P$^o$--3p $^3$P)}} \\*
4601.478       & V5   & 2 A & 104.9 & 17.3 & 0.0263 & 358.9 &   6.7 &  21.2 &   3.7 & \nodata \\
4613.868       & V5   & 2 A &  59.1 & 19.6 & 0.0177 & 414.9 &   7.7 &  24.6 &   4.3 & \nodata \\
4621.393       & V5   & 3 A & 103.9 & 18.0 & 0.0264 & 454.3 &   8.5 &  26.9 &   4.7 & \nodata \\
4630.539       & V5   & 2 A & 217.7 & 16.8 & 0.0805 & 368.6 &   6.9 &  21.8 &   3.8 & \nodata \\
4643.086       & V5   & 1 A & 140.5 & 17.0 & 0.0344 & 473.7 &   8.8 &  28.0 &   4.9 & \nodata \\*
Sum:           & V5   & \nodata & \nodata & \nodata & 0.1858 & 398.5 &  7.4 &  23.6 &   4.1 & \nodata \\
\multicolumn{11}{l}{\textbf{V15 (3p $^1$P--3d $^1$D$^o$)}} \\*
4447.030       & V15  & 2 A &  19.7 & 16.5 & 0.0039 & 2.7 & 2.7 & 6.2 & \nodata & \nodata \\
\multicolumn{11}{l}{\textbf{V20 (3p $^3$D--3d $^3$D$^o$)}} \\*
4774.244       & V20  & 1 A &  15.0 & 13.5 & 0.0022 &   97.9 &   2.8 & 117.6 &  2.3 & \nodata \\
4779.723       & V20  & 2 A &  88.9 & 17.8 & 0.0179 &  265.8 &   7.5 & 319.5 &  6.2 & \nodata \\
4781.190       & V20  & 6 A &   7.9 & 28.9 & 0.0013 &   55.8 &   1.6 &  67.1 &  1.3 & \nodata \\
4788.137       & V20  & 2 A & 101.9 & 15.6 & 0.0195 &  187.7 &   5.3 & 225.6 &  4.4 & \nodata \\
4803.286       & V20  & 4 A &  76.5 & 16.9 & 0.0261 &  141.0 &   4.0 & 169.5 &  3.3 & \nodata \\*
Sum:           & V20  & \nodata & \nodata & \nodata & 0.0670 &  166.6 &   4.7 & 200.3 &  3.9 & \nodata \\
\multicolumn{11}{l}{\textbf{V21 (3p $^3$D--3d $^3$P$^o$)}} \\*
4459.937       & V21  & 6 A &  14.3 & 17.2 & 0.0025 & \nodata  &  20.7 & 218.7 &  8.7 & \nodata \\
4477.682       & V21  & 4 A &  29.9 & 25.1 & 0.0061 & \nodata  &  22.5 & 238.5 &  9.5 & \nodata \\
4507.560       & V21  & 5 B &  25.8 & 13.1 & 0.0051 & \nodata  &  10.2 & 107.5 &  4.3 & \nodata \\*
Sum:           & V21  & \nodata & \nodata & \nodata & 0.0137 & \nodata  &  15.3 & 162.2 &  6.5 & \nodata \\
\multicolumn{10}{l}{\textbf{V24 (3p $^3$S--3d$^3$P$^o$)}} \\*
4994.371       & V24  & \nodata & 146.2 & 18.6 & 0.0423 &  175.4 &   9.0 & 169.2 &  6.8 & \nodata \\
\multicolumn{11}{l}{\textbf{V28 (3p $^3$P--3d$^3$D$^o$)}} \\*
5927.820       & V28  & 2 A & 113.8 & 17.4 & 0.0191 &  278.4 &   7.9 & 345.0 &  6.7 & \nodata \\
5931.790       & V28  & 2 A & 156.7 & 16.7 & 0.0273 &  177.0 &   5.0 & 219.3 &  4.3 & \nodata \\
5940.240       & V28  & 1 A &  63.0 & 18.2 & 0.0139 &  270.5 &   7.7 & 353.1 &  6.5 & \nodata \\
5941.650       & V28  & 1 A & 137.3 & 16.0 & 0.0315 &  109.6 &   3.1 & 135.7 &  2.6 & \nodata \\*
5952.390       & V28  & 3 A &  33.7 & 13.8 & 0.0052 &  101.6 &   2.9 & 125.9 &  2.4 & \nodata \\*
Sum:           & V28  & \nodata & \nodata & \nodata & 0.0970 &  158.2 &   4.5 & 196.1 &  3.8 & \nodata \\
\multicolumn{11}{l}{\textbf{V29 (3p $^3$P--3d $^3$P$^o$)}} \\*
5452.071       & V29  & 3 A &  25.9 & 20.2 & 0.0042 &   91.1 &   4.7 & 135.4 &  5.4 & \nodata \\
5462.581       & V29  & 2 A &  36.4 & 26.2 & 0.0045 &  130.7 &   6.8 & 194.4 &  7.8 & \nodata \\
5478.086       & V29  & 2 A &  19.6 & 16.6 & 0.0030 &   52.4 &   2.7 &  77.9 &  3.1 & \nodata \\
5480.050       & V29  & 1 A &  34.9 & 17.0 & 0.0056 &   97.9 &   5.1 & 145.6 &  5.8 & \nodata \\
5495.655       & V29  & 3 A &  87.5 & 19.1 & 0.0148 &   86.6 &   4.5 & 128.8 &  5.2 & \nodata \\*    
Sum:           & V29  & \nodata & \nodata & \nodata & 0.0321 &   87.8 &   4.5 & 130.5 &  5.2 & \nodata \\
\multicolumn{11}{l}{\textbf{V30 (3p $^3$P--4s$^3$P$^o$)}} \\*    
3829.795       & V30  & 3 A &  34.2 & 21.3 & 0.0232 &  292.2 &  95.0 & 176.2 & 59.7 & \nodata \\
3842.187       & V30  & 2 A &  33.6 & 19.1 & 0.0126 &  199.6 &  64.9 & 120.4 & 40.8 & \nodata \\
3855.096       & V30  & 2 A &  29.9 & 29.6 & 0.0098 &  155.6 &  50.6 &  93.8 & 31.8 & \nodata \\*
Sum:           & V30  & \nodata & \nodata & \nodata & 0.0456 &  221.9 &  72.2 & 133.8 & 45.3 & \nodata \\
\multicolumn{11}{l}{\textbf{V31 (3p $^1$D--3d $^1$F$^o$)}} \\*
6610.560       & V31  & 2 A &  11.7 & 25.0 & 0.0028 &   11.8 &  11.8 &  10.9 &\nodata & \nodata \\
\multicolumn{11}{l}{\textbf{V36 (3d $^3$F$^o$--4p $^3$D)}} \\*
6167.750       & V36  & 6 D &  15.9 & 22.1 & 0.0025 &    2.0 &   1.7 &   1.2 &  1.0 & \nodata \\
6170.160       & V36  & 1 A &   8.7 & 15.5 & 0.0009 &    1.5 &   1.3 &   0.9 &  0.8 & \nodata \\
6173.310       & V36  & 1 A &  20.1 & 13.6 & 0.0023 &    2.6 &   2.3 &   1.6 &  1.3 & \nodata \\*
Sum:           & V36  & \nodata & \nodata & \nodata & 0.0057 &    2.2 &   1.8 &   1.2 &  1.1 & \nodata \\
\multicolumn{11}{l}{\textbf{V54 (3d $^3$P$^o$--4p $^3$S)}} \\* 
6809.970       & V54  & 4 A &  12.6 & 24.6 & 0.0021 &   38.2 &   8.4 &  13.9 &  2.6 & \nodata \\
6834.090       & V54  & 5 B &   6.5 & 31.9 & 0.0016 &   48.8 &  10.8 &  17.8 &  3.3 & \nodata \\*
Sum:           & V54  & \nodata & \nodata & \nodata & 0.0037 &   47.5 &  10.5 &  17.3 &  3.2 & \nodata \\
\multicolumn{11}{l}{\textbf{V61.15 (4p $^3$D--5s $^3$P$^o$)}}\\*
8676.090       & V61.15 & 5 A & 4.9 & 32.3 & 0.0012 &  \nodata &   1.7 & \nodata &\nodata & \nodata \\
\enddata
\end{deluxetable*}

\addtocounter{table}{-1}
\begin{deluxetable*}{lccrrcrr@{\extracolsep{0.2pc}}r@{\extracolsep{0pc}}rc}
\tablecolumns{11}
\tablewidth{6.9in}
\tabletypesize{\footnotesize}
\tablecaption{(continued)}
\tablehead{
 & & & & & & \multicolumn{4}{c}{N$^{+2}$/H$^+$} & \\
 & & & & & & \multicolumn{4}{c}{($10^{-4}$)} & \\
\multicolumn{1}{c}{Line(s)} & & & & \multicolumn{1}{c}{FWHM} & I(\lam)/I(H$\beta$) & \multicolumn{2}{c}{EV90\tablenotemark{1}} & \multicolumn{2}{c}{KS02\tablenotemark{2}} & \\ \cline{7-8}\cline{9-10}
\multicolumn{1}{c}{(\AA)} & \colhead{Mult} & \colhead{IDI/Rank} & \multicolumn{1}{c}{S/N} & \multicolumn{1}{c}{(km/s)} & (I(H$\beta$)=100) & \multicolumn{1}{c}{Case A} &
\multicolumn{1}{c}{Case B} & \multicolumn{1}{c}{Case A} & \multicolumn{1}{c}{Case B} & Notes 
}
\startdata
\multicolumn{11}{l}{\textbf{3d-4f}} \\*
4035.081       & V39a & 3 A &  16.6 & 37.9 & \textit{0.0068}  &   0.8 &    0.8 & \nodata & \nodata & \nodata \\
4041.310       & V39b & 3 B &  30.2 & 24.9 & \textit{0.0119}  &   0.8 &    0.8 & \nodata & \nodata & \tablenotemark{4} \\
4043.532       & V39a & 4 A &   8.7 & 24.3 & 0.0041 &    0.4 &   0.4 & \nodata &\nodata & \nodata \\
4176.159       & V43a & 3 A &  13.0 & 27.3 & 0.0064  &   0.9 &   0.9 & \nodata &\nodata & \nodata \\
4236.927       & V48a & 2 A &  18.0 & 14.3 & 0.0030  &   0.7 &   0.7 & \nodata &\nodata & \nodata \\
4433.475       & V55b & 6   &  11.9 & 23.6 & 0.0028  &   2.6 &   2.5 & \nodata &\nodata & \nodata \\
4530.410       & V58b & 3 A &  21.6 & 16.7 & 0.0042  &   0.4 &   0.4 & \nodata &\nodata & \nodata \\
4552.522       & V58b & 3 A &  11.3 & 22.1 & 0.0026  &   1.0 &  \nodata& \nodata &\nodata &\tablenotemark{5} \\
4678.135       & V61b & 2 A &   5.8 & 24.3 & 0.0014  &   0.5 &   0.5 & \nodata &\nodata & \nodata \\*
Sum:     & 3d--4f & \nodata & \nodata & \nodata & \nodata & 0.7 &   0.7 & \nodata &\nodata &\tablenotemark{6} \\
\enddata

\tablenotetext{1}{From \citet{EV90}}
\tablenotetext{2}{From \citet{KS02}}
\tablenotetext{3}{Alternate ID in Paper I, \ion{O}{2} \lam5045.119
(IDI=2) unlikely, \lam5045.099 is strongest line in multiplet.}

\tablenotetext{4}{\ion{O}{2} 3d $^4$F$_{5/2}$--4f F $^2$[2]$^o_{5/2}$
\lam4041.278 (V50c) had smaller IDI=2 but contributes only slightly
to \ion{N}{2} 3d $^3$F$^o_4$--4f G $^2$[9/2]$_5$ \lam4041.310 with IDI=3.}

\tablenotetext{5}{Multiplet effective recombination coefficient taken
from \citet{PPB91}.  Assumed branching ratio of 1.  Resulting line
effective recombination coefficient agrees well with value calculated
from results of \citet{L00}.}

\tablenotetext{6}{Total intensity used for opacity case A is
I(3d--4f)=0.0432 (including \lam4552.522) and for case
B I(3d--4f)=0.0406 (not including \lam4552.522) where I(H$\beta$)=100.}

\end{deluxetable*}

\begin{deluxetable*}{lccrccrrrc}
\tablecolumns{10}
\tablewidth{6.25in}
\tabletypesize{\footnotesize}
\tablecaption{O$^{+2}$/H$^+$ Abundances From Recombination Lines. \label{tab8}}
\tablehead{
 & & & & & & \multicolumn{3}{c}{O$^{+2}$/H$^+$} & \\
\multicolumn{1}{c}{Line(s)} & & & &\multicolumn{1}{c}{FWHM} & \colhead{I(\lam)/I(H$\beta$)} & \multicolumn{3}{c}{($10^{-4}$)} & \\ \cline{7-9} 
\multicolumn{1}{c}{(\AA)} & \colhead{Mult} & \colhead{IDI/Rank} & \multicolumn{1}{c}{S/N} & \multicolumn{1}{c}{(km/s)} & (I(H$\beta$)=100) & \multicolumn{1}{c}{Case A} &
\multicolumn{1}{c}{Case B} & \multicolumn{1}{c}{Case C} & Notes
}
\startdata
\multicolumn{10}{l}{\textbf{V1 (3s $^4$P--3p $^4$D$^o$)}} \\
4638.856       & V1   & 2 A & 125.4 & 18.4 & \textit{0.0196} &   2.0 &   1.9 & \nodata &\tablenotemark{1} \\
4641.810       & V1   & 1 A & 156.1 & 15.1 & \textit{0.0422} &   1.7 &   1.7 & \nodata & \nodata \\
4649.135       & V1   & 1 A & 210.8 & 15.6 & 0.0670 &   1.4 &   1.4 & \nodata & \nodata \\
4650.838       & V1   & 1 A &  99.3 & 15.5 & 0.0217 &   2.2 &   2.1 & \nodata & \nodata \\
4661.633       & V1   & 1 A & 115.0 & 15.6 & 0.0231 &   1.9 &   1.8 & \nodata & \nodata \\
4673.733       & V1   & 1 A &  22.8 & 16.4 & 0.0041 &   2.1 &   2.0 & \nodata & \nodata \\
4676.235       & V1   & 1 A &  77.3 & 16.2 & 0.0158 &   1.5 &   1.5 & \nodata & \nodata \\
4696.353       & V1   & 1 A &  14.4 & 14.7 & 0.0027 &   2.3 &   2.2 & \nodata & \nodata \\*
Sum:           & V1   & \nodata & \nodata & \nodata & 0.1968 &   1.7 &   1.6 & \nodata & \nodata \\
\multicolumn{10}{l}{\textbf{V2 (3s $^4$P--3p $^4$P$^o$)}} \\
4317.139       & V2   & 1 A &  57.8 & 16.4 & 0.0159 &   3.0 &   2.2 & \nodata & \nodata \\
4336.859       & V2   & 1 A &  12.2 & 11.5 & 0.0037 &   2.2 &   1.6 & \nodata & \nodata \\
4345.560       & V2   & 1 A &  39.6 & 15.8 & 0.0192 &   3.7 &   2.6 & \nodata & \nodata \\
4349.426       & V2   & 0 A &  77.3 & 18.6 & 0.0330 &   2.5 &   1.8 & \nodata & \nodata \\
4366.895       & V2   & 1 A &  53.0 & 13.4 & 0.0136 &   2.4 &   1.7 & \nodata & \nodata \\*
Sum:           & V2   & \nodata & \nodata & \nodata & 0.0854 &   2.8 &   2.0 & \nodata & \nodata \\
\multicolumn{10}{l}{\textbf{V5 (3s $^2$P--3p $^2$D$^o$)}} \\
4414.898       & V5   & 1 A & 113.7 & 15.0 & 0.0303 &   6.0 &\nodata  &   1.0 & \nodata \\
4416.975       & V5   & 3 A &  92.2 & 15.5 & 0.0197 &   7.0 &\nodata  &   1.1 & \nodata \\*
Sum:           & V5   & \nodata & \nodata & \nodata & 0.0500 &   6.4 &\nodata  &   1.0 & \nodata \\
\multicolumn{10}{l}{\textbf{V10 (3p $^4$D$^o$--3d $^4$F)}} \\
4069.623       & V10  & 1 A & \nodata & 18.9 & 0.0204 &    2.1 &\nodata  & \nodata & \nodata \\
4069.882       & V10  & 1 A & \nodata & 13.0 & 0.0199 &    1.3 &\nodata  & \nodata & \nodata \\
4072.153       & V10  & 1 A & 100.3 & 15.2 & 0.0327 &    1.4 &\nodata  & \nodata & \nodata \\
4075.862       & V10  & 2 A & \nodata & 17.0 & 0.0442 &    1.3 &\nodata  & \nodata & \nodata \\
4078.842       & V10  & 2 A &  23.0 & 16.3 & 0.0056 &    1.6 &\nodata  & \nodata & \nodata \\
4085.112       & V10  & 1 A &  22.5 & 16.7 & 0.0076 &    1.7 &\nodata  & \nodata & \nodata \\
4092.929       & V10  & 1 A &  16.8 & 10.8 & 0.0032 &    1.0 &\nodata  & \nodata & \nodata \\*
Sum:           & V10  & \nodata & \nodata & \nodata & 0.1336 &    1.4 &\nodata  & \nodata & \nodata \\
\multicolumn{10}{l}{\textbf{V11 (3p $^4$D$^o$--3d $^4$P)}} \\*
3907.455       & V11  & 3 A &  13.5 & 21.9 & 0.0035 &   99.3 &   3.6 &   3.4 & \nodata \\ 
\multicolumn{10}{l}{\textbf{V12 (3p $^4$D$^o$--3d $^4$D)}} \\*
3882.194       & V12  & 3 A &  27.4 & 13.7 & 0.0063 &    1.9 &   1.9 &   1.0 & \nodata \\
3883.137       & V12  & 4 A &  10.3 & 23.7 & 0.0027 &   30.5 &  24.7 &   3.4 & \nodata \\* 
Sum:           & V12  & \nodata & \nodata & \nodata & 0.0090 &    2.7 &   2.6 &   1.3 & \nodata \\
\multicolumn{10}{l}{\textbf{V15 (3s$^\prime$ $^2$D--3p$^\prime$ $^2$F$^o$)}} \\*
4590.974       & V15  & 2 A &  73.2 & 16.5 & 0.0143 &    7.9 & \nodata & \nodata & \nodata \\     
4596.176       & V15  & 2 A &  37.8 & 15.5 & 0.0094 &    7.4 & \nodata & \nodata & \nodata \\*            
Sum:           & V15  & \nodata & \nodata & \nodata & 0.0237 &    7.7 & \nodata & \nodata & \nodata \\
\multicolumn{10}{l}{\textbf{V16 (3s$^\prime$ $^2$D--3p$^\prime$ $^2$D$^o$)}} \\*
4351.260       & V16  & 5 A &  16.5 & 19.2 & 0.0080 &    4.6 & \nodata & \nodata & \nodata \\
\multicolumn{10}{l}{\textbf{V19 (3p $^4$P$^o$--4d $^4$P)}} \\*
4121.462       & V19  & 2 A & \nodata & 17.5 & 0.0166 &  106.5 &   6.1 &   5.7 & \nodata \\
4132.800       & V19  & 3 A &  34.8 & 11.5 & 0.0083 &   29.9 &   1.6 &   1.5 & \nodata \\
4153.298       & V19  & 3 A &  75.7 & 15.1 & 0.0184 &   55.5 &   2.4 &   2.3 & \nodata \\*
Sum:           & V19  & \nodata & \nodata & \nodata & 0.0433 &   56.6 &   2.8 &   2.6 & \nodata \\
\multicolumn{10}{l}{\textbf{V20 (3p $^4$P$^o$--3d$^4$D)}} \\*    
4104.990       & V20  & 1 A &  10.4 & 14.2 & 0.0046 &    5.4 &   0.9 &   0.8 & \nodata \\ 
4110.786       & V20  & 1 A &  19.2 & 15.0 & 0.0075 &   24.0 &   3.2 &   3.0 & \nodata \\
4119.216       & V20  & 1 A &  61.2 & 17.3 & 0.0219 &    2.6 &   2.5 &   1.4 & \nodata \\ 
4120.278       & V20  & 2 A &  \nodata & 17.4 & 0.0068 &   12.0 &   8.8 &   1.2 & \nodata \\* 
Sum:           & V20  & \nodata & \nodata & \nodata & 0.0408 &    4.0 &   2.4 &   1.4 & \nodata \\
\multicolumn{10}{l}{\textbf{V25 (3p $^2$D$^o$--3d $^2$F)}} \\*
4705.346       & V25  & 4 A &  61.4 & 14.6 & 0.0183 &   17.1 &  16.6 &   1.0 & \nodata \\ 
\multicolumn{10}{l}{\textbf{V28 (3p $^4$S$^o$--3d$^4$P)}} \\* 
4890.856       & V28  & 4 A &  50.5 & 33.5 & 0.0137 &  208.6 &  12.1 &  11.2 & \nodata \\
4906.830       & V28  & 3 A &  22.5 & 13.1 & 0.0042 &   34.5 &   1.7 &   1.6 & \nodata \\ 
4924.529       & V28  & 1 A &  25.1 & 16.4 & 0.0089 &   49.0 &   2.1 &   2.0 & \nodata \\* 
Sum:           & V28  & \nodata & \nodata & \nodata & 0.0268 &   72.6 &   3.5 &   3.2 & \nodata \\
\multicolumn{10}{l}{\textbf{V36 (3p$^\prime$ $^4$S$^o$--3d$^\prime$ $^4$P)}} \\ *
4185.439       & V36  & 1 A &  22.9 & 16.2 & 0.0081 &    3.2 &\nodata  &\nodata  & \nodata \\
4189.788       & V36  & 2 A &  34.0 & 21.2 & 0.0124 &    3.8 &\nodata  &\nodata  & \nodata \\* 
Sum:           & V36  & \nodata & \nodata & \nodata & 0.0205 &    3.5 &\nodata  &\nodata  & \nodata \\ 
\enddata
\end{deluxetable*}

\addtocounter{table}{-1}
\begin{deluxetable*}{lccrccrrrc}
\tablecolumns{10}
\tablewidth{6.25in}
\tabletypesize{\footnotesize}
\tablecaption{(continued)}
\tablehead{
 & & & & & & \multicolumn{3}{c}{O$^{+2}$/H$^+$} & \\
\multicolumn{1}{c}{Line(s)} & & & &\multicolumn{1}{c}{FWHM} & \colhead{I(\lam)/I(H$\beta$)} & \multicolumn{3}{c}{($10^{-4}$)} & \\ \cline{7-9} 
\multicolumn{1}{c}{(\AA)} & \colhead{Mult} & \colhead{IDI/Rank} & \multicolumn{1}{c}{S/N} & \multicolumn{1}{c}{(km/s)} & (I(H$\beta$)=100) & \multicolumn{1}{c}{Case A} &
\multicolumn{1}{c}{Case B} & \multicolumn{1}{c}{Case C} & Notes
}
\startdata
\multicolumn{10}{l}{\textbf{3d-4f}} \\*
4087.153       & V48c &  2 A & 24.5 & 14.5 & 0.0045 &    1.5 & \nodata & \nodata & \nodata \\
4089.288       & V48a &  2 A & 37.9 & 13.3 & 0.0114 &    1.0 & \nodata & \nodata & \nodata \\
4095.644       & V48c &  2 A & 19.4 & 10.5 & 0.0042 &    1.9 & \nodata & \nodata & \nodata \\
4097.257       & V48b &  2 A & 42.8 & 14.9 & \textit{0.0095} &    1.4 & \nodata & \nodata & \nodata \\
4275.551       & V67a &  2 A & 13.5 & 33.4 & 0.0065 &    1.1 & \nodata & \nodata & \nodata \\
4277.427       & V67c &  3 A & 10.7 & 24.9 & \textit{0.0033} &    2.3 & \nodata & \nodata & \nodata \\ 
4281.313       & V53b &  3 B &  9.1 & 20.9 & 0.0019 &    3.3 & \nodata & \nodata & \nodata \\
4282.961       & V67c &  2 A & 13.4 & 16.1 & 0.0026 &    2.3 & \nodata & \nodata & \nodata \\
4285.684       & V78b &  2 A & 16.5 & 13.2 & 0.0031 &    1.5 & \nodata & \nodata & \nodata \\
4294.782       & V53b &  2 A & 21.1 & 27.8 & 0.0059 &    2.3 & \nodata & \nodata & \nodata \\
4303.823       & V53a &  2 A &  7.9 & 14.6 & 0.0066 &    1.9 & \nodata & \nodata & \nodata \\	
4307.232       & V53b &  3 A & 45.3 & 19.5 & 0.0119 &   10.4 & \nodata & \nodata & \tablenotemark{2} \\
4466.435       & V86b &  3 A & 32.4 & 25.1 & 0.0029 &    2.9 & \nodata & \nodata & \nodata \\
4489.461       & V86b &  3 A &  7.0 & 23.0 & 0.0007 &    1.0 & \nodata & \nodata & \nodata \\
4602.129       & V92b &  2 A & 12.0 & 12.6 & 0.0022 &    1.2 & \nodata & \nodata & \nodata \\
4609.436       & V92a &  2 A & 23.7 & 15.3 & 0.0060 &    1.3 & \nodata & \nodata & \nodata \\ 
Sum:           & 3d--4f & \nodata & \nodata & \nodata  &  0.0832  &    1.7 & \nodata & \nodata & \nodata \\
\enddata

\tablenotetext{1}{Although \ion{C}{2} 4p$^2$P$^o_{3/2}$--6d
$^2$D$_{5/2}$ \lam4638.916 (V12.01) not listed as ID in Table 3 of Paper I, it likely contributes given intensity of companion line at \lam4637.630 and predicted intensity: I(\lam4638.916)=0.0101 (I(H$\beta$)=100).}

\tablenotetext{2}{Possibly contaminated by \ion{C}{2} 4p $^2$P$^o_{3/2}$--7s $^2$S$_{1/2}$ \lam4307.580 (V12.02).  Wavelength difference with observed line too large for EMILI to assign a IDI value.} 
\end{deluxetable*}

\begin{deluxetable*}{lccrrrrrc}
\tablecolumns{9}
\tablewidth{6.5in}
\tabletypesize{\footnotesize}
\tablecaption{Other Ionic Abundances from Recombination Lines. \label{tab9}}
\tablehead{
\multicolumn{1}{c}{Line(s)} & & & & \multicolumn{1}{c}{FWHM} & \multicolumn{1}{c}{I($\lambda$)/I(H$\beta$)} & \multicolumn{2}{c}{N$^{+i}$/H$^+$} & \\ \cline{7-8}
\multicolumn{1}{c}{(\AA)} & \colhead{Mult} & \colhead{IDI/Rank} & \multicolumn{1}{c}{S/N} & \multicolumn{1}{c}{(km/s)} & \multicolumn{1}{c}{I(H$\beta$=100))} & \multicolumn{1}{c}{Case A} & \multicolumn{1}{c}{Case B} & \colhead{Notes}
}
\startdata \tableline
\multicolumn{9}{c}{\textbf{He$^+$/H$^+$} (10$^{-2}$)} \\* \tableline
5875.615      & V11 & 4 A & 5395.0 & 25.9 & 13.6746 &   9.5 & \nodata & \nodata \\
4471.474      & V14 & 4 A & 1935.0 & 23.7 &  4.4921 &   9.0 & \nodata & \nodata \\
6678.152      & V46 & 2 A & 2346.0 & 21.6 &  3.8721 &  10.3 &  10.1 & \nodata \\*
Average  & \nodata & \nodata & \nodata & \nodata & 22.0388 & 9.5 & 9.5 & \tablenotemark{1} \\ \tableline
\multicolumn{9}{c}{\textbf{N$^{+}$/H$^+$} (10$^{-4}$)} \\* \tableline
\multicolumn{9}{l}{\textbf{V1 (3s $^4$P -- 3p $^4$D$^o$)}} \\*
8680.282      & V1  & 1 A &  396.1 & 56.8 &  0.0385 &  11.9	&  11.5 & \nodata \\
8683.403      & V1  & 3 A &  382.0 & 57.6 &  0.0417 &  24.7	&  23.9 & \nodata \\
8686.149      & V1  & 2 A & \nodata & 52.3 &  0.0220 &  32.7	&  31.7 & \nodata \\
8703.247      & V1  & 1 A &  254.2 & 54.0 &  0.0268 &  39.9	&  38.7 & \nodata \\
8711.703      & V1  & 1 A &  234.7 & 52.8 &  0.0273 &  31.8	&  30.8 & \nodata \\
8718.837      & V1  & 2 A &  143.9 & 53.4 &  0.0135 &  18.6	&  18.1 & \nodata \\*
Sum:          & V1  & \nodata & \nodata & \nodata & 0.1698 &  21.6 &  21.0 & \nodata \\
\multicolumn{9}{l}{\textbf{V2 (3s $^4$P -- 3p $^4$P$^o$)}} \\*
8184.862      & V2  & 7 D &  168.0 & 20.8 &  0.0152 &  29.7	&  25.3 & \nodata \\
8200.357      & V2  & 2 A &  138.6 & 51.5 &  0.0131 & 138.8	& 118.0 & \nodata \\
8216.336      & V2  & 5 B &  304.1 & 62.2 &  0.0598 &  50.4	&  42.9 & \nodata \\
8223.128      & V2  & 1 A &  347.8 & 50.4 &  0.0623 & 132.4	& 112.7 & \nodata \\
8242.389      & V2  & 4 B & \nodata & 53.9 &  0.0551 &	108.8	&  92.6 & \nodata \\*
Sum:          & V2  & \nodata & \nodata & \nodata & 0.2055 &  74.2 &  63.2 & \nodata \\
\multicolumn{9}{l}{\textbf{V3 (3s $^4$P -- 3p $^4$S$^o$)}} \\*
7423.641      & V3  & 1 A &  119.4 & 58.3 & 0.0156 & 157.5 &  52.5 & \nodata \\
7442.298      & V3  & 1 A &  219.9 & 57.7 & 0.0363 & 184.0 &  61.3 & \nodata \\
7468.312      & V3  & 3 A & \nodata & 59.2 & 0.0583 & 198.3 &  66.1 & \nodata \\*
Sum:          & V3  & \nodata & \nodata & \nodata & 0.1102 & 186.7 &  62.2 & \nodata \\ \tableline
%
\multicolumn{9}{c}{\textbf{O$^{+}$/H$^+$} (10$^{-4}$)} \\* \tableline
\multicolumn{9}{l}{\textbf{V1 (3s $^5$S$^o$ -- 3p $^5$P)}} \\*
7771.944     & V1   & 2 A & 157.4 & 38.4 & 0.0352 &   3.6 & \nodata & \nodata \\
7774.166     & V1   & 1 A &  148.9 & 27.1 & 0.0215 &   3.1 & \nodata & \nodata \\
7775.387     & V1   & 1 A &  148.9 & 28.6 & 0.0130 &   3.1 & \nodata & \nodata \\*
Sum:         & V1   & \nodata & \nodata & \nodata &  0.0697 &   3.3 & \nodata & \nodata \\
\multicolumn{9}{l}{\textbf{V4 (3s $^3$S$^o$ -- 3p $^2$P)}} \\*
8446.247,.359,.758 & V4 & 7 B & \nodata & 68.1 & 1.1419 & 428.8 & 98.8 & \nodata \\ 
\multicolumn{9}{l}{\textbf{V8 (3p $^5$P -- 3d $^5$D$^o$)}} \\*
9260.806,.848,.937 & V8 & 1 A & \nodata & 43.8 & 0.0047 &   3.0 & \nodata & \nodata \\*
9262.582,.670,.776 & V8 & 1 A & \nodata & 43.7 & 0.0097 &   3.8 & \nodata & \nodata \\*
9265.826,.932,6.005 & V8 & 1 A & 99.0   & 27.8 & 0.0125 &   3.5 & \nodata & \nodata \\*
Sum:         & V8 & \nodata & \nodata & \nodata &  0.0269 &   3.5 & \nodata & \nodata \\ \tableline
\multicolumn{9}{c}{\textbf{Ne$^{+2}$/H$^+$ (10$^{-5}$)}} \\* \tableline
\multicolumn{9}{l}{\textbf{3d--4f}} \\*
4219.745     & V52a & 2 A & 8.0 & 19.9 & 0.0015 & 2.8 & \nodata & \nodata \\
4391.991     & V55e & 3 A & 7.4 & 15.8 & 0.0022 & 2.3 & \nodata & \tablenotemark{2} \\*
Average:     & 3d--4f & \nodata & \nodata & \nodata & 0.0037 & 2.4 & \nodata & \nodata \\
%
\enddata

\tablenotetext{1}{Case B average includes Case A \lamm4471.474,5875.615 values.}

\tablenotetext{2}{Ignoring \ion{Ne}{2} 3d $^4$F$_{9/2}$--4f
$^2$[5]$^o_{9/2}$ \lam4391.995 (V55e) as was done by \citet{L00}.}

\end{deluxetable*}

\begin{figure*}
\epsscale{0.7}
\plotone{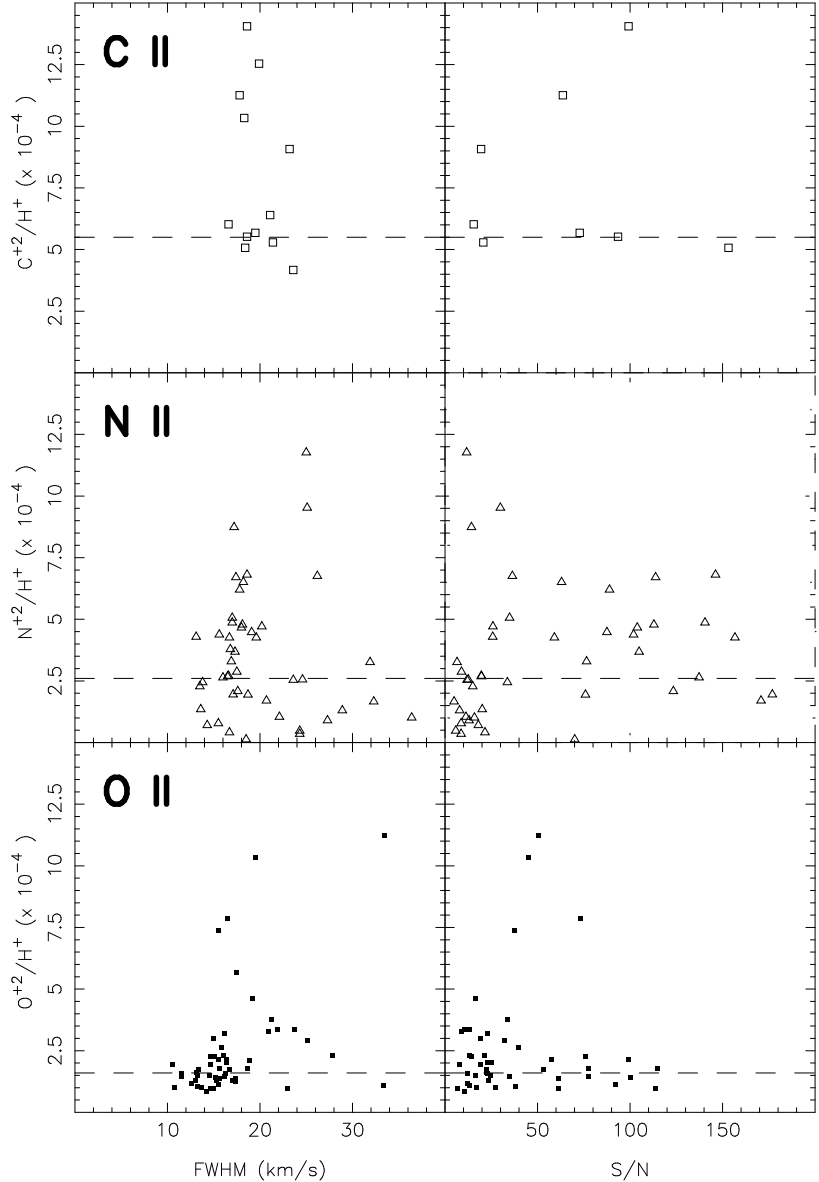}
\epsscale{1.0}
\caption{\ion{C}{2}, \ion{N}{2}, \ion{O}{2} line-derived abundances
versus FWHM and S/N from Tables~\ref{tab6}--\ref{tab8}.  The dashed
lines indicated the final assumed abundance for those
ions. \label{f9}}
\end{figure*}

Although permitted lines from numerous ionic species are found within
our spectra, we concentrate mainly on ions represented by both
collisionally excited lines and radiative recombination lines selected
by the above procedure: N$^+$, O$^+$, O$^{+2}$, and Ne$^{+2}$, and for
other ions with prominent recombination line spectra: He$^+$,
C$^{+2}$, and N$^{+2}$.  Several works have discussed the excitation
mechanisms of specific ions' purported recombination lines
\citep{G75a,G75b,G76,L95,E98,E99,L00}.  We compare our results with
these references' interpretations for their excitation sources.

\subsubsection{C II}

\citet{G76} has qualitatively shown that \ion{C}{2} \lam4267
(multiplet V6) is excited primarily by radiative recombination, a
conclusion also reached by \citet{H80} for IC 418.  The excellent
agreement seen between the \lam4267 abundance and those obtained from
the numerous high excitation, high angular momentum upper term lines
listed in Table~\ref{tab6} confirms this, as such terms are not
excitable by ground term fluorescence processes and should be populated
exclusively by radiative recombination.  

The remaining lines listed in Table~\ref{tab6} come from terms for
which \citet{G76} has calculated that continuum fluorescence plays a
dominate role in exciting in the Orion Nebula.  Continuum fluorescence
from the the 2p $^2$P$^o$ ground term via central star radiation in
resonance lines is capable of exciting higher $^2$S and $^2$D terms
directly (e.g.\ the upper terms of multiplets V3, V4, V10.03 and
V12.01) or $^2$P$^o$ terms indirectly following cascades from excited
$^2$S and $^2$D terms (multiplets V2 and V5).  The abundance pattern
recorded for these lines in Table~\ref{tab6} agrees generally with
such a scenario, with the larger abundance from multiplet V4
reflecting the lack of alternate optical cascade paths for its 3s
$^2$S upper term and its direct excitation through a resonance line.
\citet{G76} predicts a recombination-only intensity of the multiplet
V4 line \lam3921 around a tenth the observed strength, an amount
which is coincidentally sufficient to align its abundance determined
here with that of \lam4267.
 
However, the evidence for continuum fluorescence excitation among
\ion{C}{2} lines arising from purported continuum
fluorescence-enhanced terms is not corroborated by line profile
difference with those from lines thought to be excited by radiative
recombination alone.  A comparison of line profiles in Figure~\ref{f6}
between \lamm3919, 3921 (multiplet V4), \lam5892 (multiplet V5), and
\lam6260 (multiplet V10.03) which should be enhanced by continuum
fluorescence, and those of higher angular momentum and excitation
lines \lam6462 (multiplet V17.04) and \lam5342 (multiplet V17.06)
which are unaffected by continuum fluorescence, shows no remarkable
difference.  Plots of the FWHM and S/N versus calculated abundance for
lines listed in Table~\ref{tab6}, shown in Figure~\ref{f9}, shows no
discernible trend.  If continuum fluorescence was the dominate or
equal exciter for the $^2$S, $^2$P$^o$, and $^2$D upper term lines in
IC 418 with respect to radiative recombination, it would be expected
that lines coming from these terms would have profiles and FWHM
similar to [\ion{N}{2}] or [\ion{O}{2}] collisionally excited lines as
they their parent ions have a similar ionization potential as C$^+$
and such ions should be nearly co-spatial in the nebula.  In addition
the calculations of \citet{G76} also predict a fluorescence
enhancement, by factor of 2, for multiplet V3 lines which is not
reflected in the abundances determined from its constituent lines
observed here.  \citet{E98} fails to show any enhancement of either
multiplet V2 or V4 in the Orion Nebula, nor does \citet{E99} show any
enhancement for multiplet V4 in M8.  This suggests that the excitation
mechanism varies in efficiency from object to object.

From the above, we conclude that while the strengths of multiplets V2,
V4, V5, V10.02, and V12.03 appear to have been enhanced, the mechanism
for this is uncertain.  These multiplets therefore were excluded from
the calculation of the final $C^{+2}$/H$^+$ abundance.  For the
remaining lines, each case-sensitive multiplet yields an abundance in
case B which corresponds to case-insensitive \lam4267.  We therefore
adopt the summed abundances determined from multiplets V3, V6, and
high excitation lines with upper term angular momentum $L>2$.

\subsubsection{N II}

Abundances for \ion{N}{2} permitted lines were computed under two
opacity cases with two sources of multiplet effective recombination
coefficients.  The difference in the two sources of coefficients
appears to lie in their treatment of low temperature dielectronic
recombination.  For nearly all triplet multiplets, using the
\citet{KS02} values resulted in lower abundances than those calculated
using the \citet{EV90} values, and for multiplets observed in common
with \citet{L00} (multiplets V3, V5, V20, V28), slightly better
consistency.  Thus, in accordance with comparisons between
case-sensitive and insensitive multiplets, the \citet{KS02} derived
case B abundances for all triplet multiplets are adopted here.  For
the singlet lines, \lam4437 (multiplet V15) had a smaller abundance
using the \citet{EV90} values in both opacity cases, while \lam6611
(multiplet V31) had nearly equal abundances from both sources.  This
may reflect uncertainties in their recombination coefficients, and
they are not discussed further here.

The abundance pattern that emerges from Table~\ref{tab7} fits a
scenario quantitatively proposed by \citet{G76} to explain the
\ion{N}{2} emission in the Orion Nebula, and reasonably extrapolated
to the larger group of lines observed here in IC 418.  The first
element is excitation of the 4s $^3$P$^o$ term via pumping of the
\ion{N}{2} 2p$^2$ $^3$P$_0$ - 2p4s $^3$P$^o_1$ \lam508.668\AA (and
\lam508.697\AA; Escalante 2002) resonance lines by the \ion{He}{1}
1s$^2$ $^1$S -- 1s8p$^1$P$^o$ \lam508.643\AA\ recombination line
leading to strong multiplet V30 emission.  This is followed by
subsequent cascade to enhance multiplets V3 and V5.  Starlight
continuum fluorescence excitation of the 3d $^3$P$^o$, 3d $^3$D$^o$,
and to a lesser extent the 5s $^3$P$^o$ upper terms of multiplets V20,
V21, V24, V28, V29, and V61.15 via resonance transitions from the
ground $^3$P term, supplement radiative recombination in populating
these terms.  Cascade via multiplet V24 excites multiplet V4 in turn.
All these multiplets show enhanced emission and consequent higher
abundances over multiplets better described by pure radiative
recombination case, as represented by multiplet V36, and the 3d--4f
transitions, which lack direct resonance or subsequent cascade
excitation paths. Only multiplet V54 would does not fit this paradigm.
Considerable scatter within the 3p--3d multiplet's individual line
abundances with FWHM and S/N, as depicted in Figure~\ref{f9}, agrees
with likely departures from $LS$-coupling \citep{E98,KS02}, while
3d--4f lines, lacking an alternate excitation source are weaker, have
smaller S/N, and cluster around lower abundances.  However, the
strongest members of the 3d--4f transition group: \lam4035 (multiplet
V39a) and \lam4041 (multiplet V39b), are comparable in intensity to
some 3--3 transitions, and still yield abundances consistent with
other 3d--4f transitions.

As with \ion{C}{2}, evidence for continuum or line fluorescence
excitation is not seen among the \ion{N}{2} line profiles shown in
Figure~\ref{f6}.  The profiles of \lamm3839, 3842 of multiplet V30,
lines yielding the highest abundances among all the \ion{N}{2} lines
observed here due purportedly to excitation by the \ion{He}{1} line
fluorescence mechanism, are no different than any other \ion{N}{2}
permitted lines depicted in Figure~\ref{f6}.  No distinct abundance
trend with FWHM is seen in Figure~\ref{f9}.  Comparison with a recent
\ion{N}{2} permitted line emission model of the Orion Nebula
\citep{E02}, suggests that the \ion{He}{1} line fluorescence mechanism
is necessary to generate appreciable multiplet V30 emission.  However,
the same model yields a summed intensity of multiplet V30 emission a
factor of four larger than our observed value in terms of percentage
of the total amount of emission expected from all \ion{N}{2} lines
modeled/observed in common.  The evidence for a large \ion{He}{1}
mechanism operating in IC 418 is therefore inconclusive from both the
emission model and profiles, despite the seemingly large degree of
enhancement in multiplet V30 line abundances, and the additional
excitation mechanism responsible for these lines' strong intensity is
unknown.

For the final N$^{+2}$/H$^+$ abundance, multiplet V30 was excluded
from the calculation due to confusion regarding its excitation source.
However, since the remaining purported continuum fluorescence-enhanced
lines show clear recombination-like profiles, they were included
assuming opacity case B.  Singlet lines were not included in this
calculation.
  
\subsubsection{O II}\label{oii}

Our deep observations reveal a rich \ion{O}{2} permitted line spectrum
for IC 418 consisting of lines from both the doublet and quartet
sequences.  The procedure used by \citet{L00} in their analysis of
their NGC 6153 spectrum was followed in that, for the 3s--3p
transitions, the $LS$-coupling based effective recombination
coefficients from \citet{S94} were used, and, for the 3p--3d and
3d--4f transitions, the \citet{L95} intermediate-coupling calculated
effective recombination coefficients were used.  The low temperature
dielectronic recombination coefficients of \citet{NS84} were used to
obtain abundances for multiplets V15, V16, and V36, are assumed to be
in opacity case A.

\citet{L00} assumed opacity case A for doublets and case B for
quartets, but our results suggest that case C, optically thick
transitions to the 2p$^3$ $^2$D$^o$ doublet ground term, yields the
most consistent abundance among all multiplets.  This would
necessitate a much larger population in the $^2$D$^o$ ground term than
would be expected under typical nebular physical conditions, although
its assumption would seem slightly more physically plausible in IC 418
(with an O$^+$-derived density of 10000 cm$^{-3}$) than it would in
the less dense NGC 6153 (3500 cm$^{-3}$; Liu et al.\ 2000), as the IC
418 density significantly exceeds the the $\sim$4000 cm$^{-3}$
critical density of the ground $^2$D$^o$ term.  Because case B and C
insensitive multiplets V10, V12, V20, and V28 yield near identical
abundance under both intermediate-coupling and $LS$-coupling
calculations (using Storey 1994, not shown in Table~\ref{tab8}) in
both cases, the choice of case B may be more likely.  However, in
order to reconcile doublet multiplets V5 and V25 abundances with the
quartets, case C must be assumed for them.  Since lines arising from
\ion{O}{2} doublets are less likely to be excited by continuum
fluorescence due to difference in multiplicity with the O$^+$ 2p$^3$
$^4$S$^o$ ground term, they are more likely to represent the effects
of purely radiative recombination (with the exception of lines from
V15, V16, and V36 discussed below), and such a reconciliation is
necessary.  Thus, in the discussion below, case C, or the highest
opacity case available, is assumed for each multiplet.

The abundance pattern in Table~\ref{tab8} fits in with the \citet{G76}
calculations that all \ion{O}{2} permitted line intensities observed
in the Orion Nebula are adequately explained by radiative
recombination alone.  Although \citet{HAF94} speculates that some
\ion{O}{2} lines in IC 418 may be susceptible to continuum
fluorescence, \citet{G76} predicts only 20\% contribution at most for
multiplet V2, and much less for other of the observed multiplets.
There is no evidence for continuum fluorescence excitation among any
of the line profiles shown in Figure~\ref{f6}.  Abundances determined
from individual lines within multiplets show good agreement, better
than what was exhibited in the \ion{N}{2} lines, validating the need
for intermediate-coupling effective recombination coefficients for the
3p--3d transitions in both \ion{N}{2} and \ion{O}{2} lines.

An inspection of Figure~\ref{f9} does show a trend in increasing
abundance with larger FWHM.  Some of the lines with higher abundances
correspond to lines from multiplets V15, V16, and V36 which are
produced primarily by low temperature dielectronic recombination.
These are discussed further in \S~\ref{dielec}.  One outlier, \lam4891
(multiplet V28), has no other obvious ID, and shows no enhancement
relative to the other lines of its multiplet in NGC 6153 \citep{L00}.
The 3d--4f line \lam4307 (multiplet 53b), another outlier, might be
contaminated by \ion{C}{2} \lam4308 (multiplet V12.02), although there
is a large wavelength disagreement, and the contribution might be
insufficient to account for the large abundance value given the
\ion{C}{2} line's likely intensity compared to other lines observed
from the $l=0$, $^2$S sequence.  Much of the remaining trend can be
ascribed to weak lines, as seen in larger scatter of abundance values
at low S/N depicted in Figure~\ref{f9}, so it is believed that no real
trend in abundance exists with FWHM.

The 3d--4f transitions, with the exception of \lam4308, show
consistent agreement among themselves, and with abundances determined
from other multiplets under their case B or C conditions.  These
transitions are immune to fluorescence excitation and are case
insensitive.  Although many are weak, the strongest ones, such as
\lam4089 (multiplet V48a), whose branching ratio is insensitive to
assumed coupling case \citep{L00}, seem to be good abundance
indicators.

Larger abundances were calculated from the intensities of multiplets
V15, V16, and V36, all of which are have large low-temperature
dielectronic recombination coefficients \citep{NS84}.  This agrees
with the trend seen by \citet{GD01a} in multiplet V15 among other PNe.
The abundances determined from these lines are not included in
the final O$^{+2}$/H$^+$ calculation and are discussed further in
\S~\ref{dielec}.

To compute the final O$^{+2}$/H$^+$ abundance, the highest opacity
case values available for all lines was used, excluding multiplets
V15, V16, and V36.

\subsubsection{Permitted Lines From Other Ions}

As in \citet{L00}, the He$^+$/H$^+$ abundance was calculated from the
average $\lambda$4471, $\lambda$5876, and $\lambda$6678 individual
line abundances, weighted 1:3:1, using the tabulated emissivites of
\citet{S96} at $T_e$=10000 K and $N_e$=10000 cm$^{-3}$.  Line
intensities were corrected for electron collisional excitations from
the He$^o$ 2s $^3$S metastable term by the formalism of \citet{KF95}
using $T_e$=9600 K and $N_e$=10000 cm$^{-3}$.  Only case A
calculations were available for \lamm4471, 5876.  Case B was assumed for
\lam6678 in the calculated average.

For the \ion{N}{1} lines, the profiles in Figure~\ref{f6} and the
abundances listed in Table~\ref{tab9} clearly show that all lines from
quartet multiplets are excited predominantly by continuum fluorescence,
as was shown to be the case in the Orion Nebula by \citet{G75b}.
Direct multiplet effective recombination coefficients are available
only for quartets multiplets in \citet{PPB91}.  Weak \ion{N}{1}
doublet lines \lam8567.735 and \lam 8629.236 (both multiplet V8) may
also be identified within our spectra.  The \lam8567 lines sits in a
gap among the OH Meinel (6-2) band P branch and but has a peculiar
profile on the CCD image which suggests that it might be
a scattered light artifact.  The \lam8629 lines is blended with an
O$_2$ $b$--$X$ (0-1) band feature but possesses a distinct nebular-like
profile in the two-dimensional spectrum, and would be the strongest
line in the multiplet.  Employing the corresponding \ion{O}{2}
transition effective recombination coefficient \citep{S94} yields
approximate, opacity case insensitive-abundances of about 2 $\times
10^{-4}$ and 4 $\times 10^{-4}$ for the \lamm8567, 8629 lines
respectively, lower than the triplet values, but still a factor of ten
times higher than the corresponding collisional line value for
N$^+$/H$^+$.

\ion{O}{1} \lam8446 (multiplet V4) has long been known to be excited
primarily by starlight continuum fluorescence through excitation via
resonance transitions to higher excitation $^3$S$^o$ and $^3$D$^o$
terms from the neutral oxygen $^3$P ground term and subsequent cascade
\citep{G75a}.  The profiles of \lam8446 (a blend of three
transitions), \lam5299, and \lam5513, all of which belong to triplet
multiplets, show similar morphology to the [\ion{O}{1}] \lamm6300,
6364 collisionally excited lines, as seen in Figure~\ref{f6}, and all
have similar FWHM, indicating common spatial origin and likely
predominate fluorescence excitation.  At least 15 other higher
excitation lines with good IDI values not listed in Table~\ref{tab9}
are visible in triplet and quintet Rydberg sequences in our spectrum.
Quintet lines are much weaker than their triplet counterparts.  The
consistency of abundances determined from them, and their line
profiles (\lam7772, \lam9266, and \lam6157 in Figure~\ref{f6}) suggest
radiative recombination as their primary excitation source.

The spectrum of IC 418 lacks large numbers of \ion{Ne}{2} permitted
lines, including the strongest 3--3 transitions seen by \citet{L00} in
NGC 6153 such as \lam3694.212 and \lam3709.621 (both multiplet V1).
Many other strong \ion{Ne}{2} permitted line multiplets reside just
blue-ward of our bandpass, however.  Within our spectrum the sole
semi-certain \ion{Ne}{2} 3--3 transition, \ion{Ne}{2} \lam3777.134
also from multiplet V1, should be weaker than either of the two other
lines mentioned above from the same multiplet, and has too large a
FWHM to be included in Table ~\ref{tab9}.  On the other hand the
strongest 3d--4f transition observed by \citet{L00}, \ion{Ne}{2}
\lam4219.745 (multiplet V52a), is observed here, as is another line,
\ion{Ne}{2} \lam4391.991 (multiplet V55e).  Both yield abundances
about a factor of two smaller than \lam3777.134, although \citet{L00}
notes that their effective recombination coefficients for 3d--4f
transitions, which were utilized here, might be in error by a factor
of 2.  Another 3d--4f line, \ion{Ne}{2} \lam4457.050 (multiplet V61d)
is also present, but, as also seen by \citet{L00} in NGC 6153, has an
anomalously high intensity, here yielding an abundance nearly two
orders of magnitude greater than any of the lines mentioned above.  As
this line also has a large FWHM, this is obviously an incorrect ID,
although no other obvious ID for it is suggested by EMILI.  The
Ne$^{+2}$/H$^+$ abundance, therefore, is drawn only from the two
3d--4f lines that appear to have the best IDs: \lam4219.745 and
\lam4391.991, although comparisons with the [\ion{Ne}{3}] forbidden
line abundances are probably not statistically significant using only
two presumed recombination lines with uncertain recombination
coefficients.

\section{Dielectronic Recombination Lines} \label{dielec}

\begin{deluxetable*}{lccrccc}
\tablecolumns{7}
\tabletypesize{\footnotesize}
\tablewidth{5in}
\tablecaption{C II, N II, O II dielectronic recombination lines. \label{tab10}}
\tablehead{
\multicolumn{1}{c}{Line(s) \lam} & & & & \multicolumn{1}{c}{FWHM} & \multicolumn{1}{c}{I($\lambda$)/I(H$\beta$)} & \\
\multicolumn{1}{c}{(\AA)} & \colhead{Mult.} & \colhead{IDI/Rank} & \multicolumn{1}{c}{S/N} & \multicolumn{1}{c}{(km/s)} & \multicolumn{1}{c}{I(H$\beta$)=100} & \colhead{Notes}}
\startdata \tableline
\multicolumn{7}{c}{\textbf{C II}} \\ \tableline
5125.208 & V12 & 1 A &  29.6 & 14.9 & 0.0058 & \nodata \\
5126.963 & V12 & 1 A &  12.7 & 16.9 & 0.0030 & \nodata \\
6779.940 & V14 & 1 A & \nodata & 17.9 & 0.0109 & \nodata \\
6780.600 & V14 & 0 A & \nodata & 17.7 & 0.0055 & \nodata \\
6783.910 & V14 & 4 A &  12.4 & 19.1 & 0.0022 & \nodata \\
6791.470 & V14 & 1 A &  51.5 & 19.2 & 0.0066 & \nodata \\
6800.680 & V14 & 0 A &  33.1 & 17.7 & 0.0050 & \nodata \\
5132.947,3.282 & V16 & 6 B & 14.9 & 39.5 & 0.0044 & \nodata \\
5145.165       & V16 & 3 B & 12.3 & 14.8 & 0.0040 & \nodata \\
5151.085       & V16 & 2 A & 16.4 & 20.5 & 0.0046 & \nodata \\
7113.040       & V20 & 5 B & 34.6 & 35.4 & 0.0052 & \nodata \\
7115.630       & V20 & 3 B & 30.8 & 21.1 & 0.0043 & \nodata \\
7119..760,.910 & V20 & 5 B & 41.5 & 42.1 & 0.0070 & \nodata \\ \tableline
\multicolumn{7}{c}{\textbf{N II}} \\* \tableline
5530.242       & V63 & 2 A & 10.8 & 19.2 & 0.0021 & \nodata \\
5535.347,.384  & V63 & 1 A & 16.9 & 26.4 & 0.0050 & \tablenotemark{1} \\
5543.471       & V63 & 1 A & 15.3 & 27.6 & 0.0045 & \nodata \\
5551.922       & V63 & 2 A &  7.5 & 12.0 & 0.0006 & \nodata \\
5179.520 & V66 & 2 A & 18.4    & 18.6 & 0.0033 & \tablenotemark{2} \\ \tableline
\multicolumn{7}{c}{\textbf{O II}} \\ \tableline
4465.408 & V94 & 1 A & 32.4 & 14.8 & 0.0059 & \tablenotemark{3} \\
4467.924 & V94 & 1 A & 22.6 & 14.7 & 0.0041 & \tablenotemark{4} \\
4469.378 & V94 & 1 A & 12.1 & 13.5 & 0.0022 & \tablenotemark{5} \\
4145.096,6.076 & V106 & 4 B & 17.3 & 19.9 & 0.0059 & \tablenotemark{6}
\enddata

\tablenotetext{1}{Possibly \ion{C}{2} 4s $^2$S$_{1/2}$--5p
$^2$P$^o_{3/2}$ \lam5535.353 (V10) but could include these two lines.
\lam5535.347 is predicted to be strongest in multiplet.}

\tablenotetext{2}{May include contribution from \ion{N}{2}
3p$^\prime$ $^5$P$^o_3$--3d$^\prime$ $^5$D$_4$ \lam5179.344 (V70),
both lines strongest in their respective multiplets.}

\tablenotetext{3}{May be blended with \ion{N}{2} 3p $^3$D$_1$--3d
$^3$P$^o_1$ \lam4465.529 (V21).  \lam4465.408 is strongest line in
multiplet.}  

\tablenotetext{4}{Alternate ID, \ion{Fe}{2} 4p y$^4$G$^o_{7/2}$--4d
e$^4$F$_{7/2}$ \lam4467.931, is unlikely.}

\tablenotetext{5}{Alternate ID, \ion{O}{2} 3d $^2$P$_{1/2}$-- 4f D
$^2$[1]$^o_{1/2}$ \lam4469.462 (V86c), is not listed in the branching
ratio table (Table 4) of Liu et al.\ (2000).  \lam4469.378 is
positioned well and is at about the right intensity relative to the
other multiplet lines.}

\tablenotetext{6}{Primary EMILI ID, \ion{Ne}{2} 4s $^2$P$_{1/2}$--5p
$^2$S$^o_{1/2}$ \lam4146.064, unlikely since lower excitation
\ion{Ne}{2} not observed at appreciable
intensites. \lamm4145.906,6.076 (former not ID'd in Paper I) together
represent together 44\% the strength of the entire multiplet.}

\end{deluxetable*}

\citet{GD01a} have reported an over-abundance of O$^{+2}$ calculated
from multiplet 15 with respect to other multiplets in numerous PNe.
We observed enhanced abundances from multiplet V15, as well as from
multiplets V16 and V36, as compared with purported one-body radiative
recombination lines.  These multiplets have large low-temperature
dielectronic recombination coefficients \citep{NS84}.  Their
abundances cannot be reconciled with the average abundance from
radiative recombination lines at any temperature over the range in
which the \citet{NS84} formalism is valid (to 60000 K).

\citet{GD01b} proposed that this is the result of enhanced
dielectronic recombination occurring at high temperatures due to an
increasing number of higher energy auto-ionizing states becoming
accessible to recombining electrons. They suggest that this occurs at
the interface between a hot central ``bubble'' and the PN shell, and
find a correlation between nebular surface brightness, a proxy for
age, and the amount of the discrepancy between abundances determined
from recombination and collisionally excited O$^{+2}$ lines.  Smaller,
younger, higher surface brightness PNe exhibit a smaller discrepancy
than do older, dimmer, larger PNe.  IC 418 would fit into this picture
as an example of a higher surface brightness PN, and thus may not be
expected to have as severe effects from dielectronic recombination as
do some other PNe.

Our deep spectra reveal numerous additional lines that from $LS$
selection rules should be produced primarily by dielectronic
recombination.  This permits us to evaluate the general importance of
dielectronic recombination.  In particular, lines from levels having a
multiplicity that differs by more than unity from that of the ground
state of the next higher stage of ionization, and which originate from
upper levels that are fed by permitted transitions from auto-ionizing
levels within a few eV of the ionization continuum are almost certain
to be strong dielectronic transitions.  \ion{C}{2}, \ion{N}{2}, and
\ion{O}{2} all have such transitions, which because of their
multiplicity or core excitation either cannot or are unlikely to be
excited by normal 2-body recombination or fluorescence processes, and
we observed and identified a number of such lines.  In
Table~\ref{tab10} we list observed transitions and intensities for
\ion{C}{2}, \ion{N}{2}, and \ion{O}{2} which based upon the Grotrian
diagrams should constitute the strongest lines from these ions due to
dielectronic recombination excitation.  They were selected based on
the same criteria as were those permitted lines listed in
Tables~\ref{tab6}--\ref{tab9}.

\begin{figure*}
\plotone{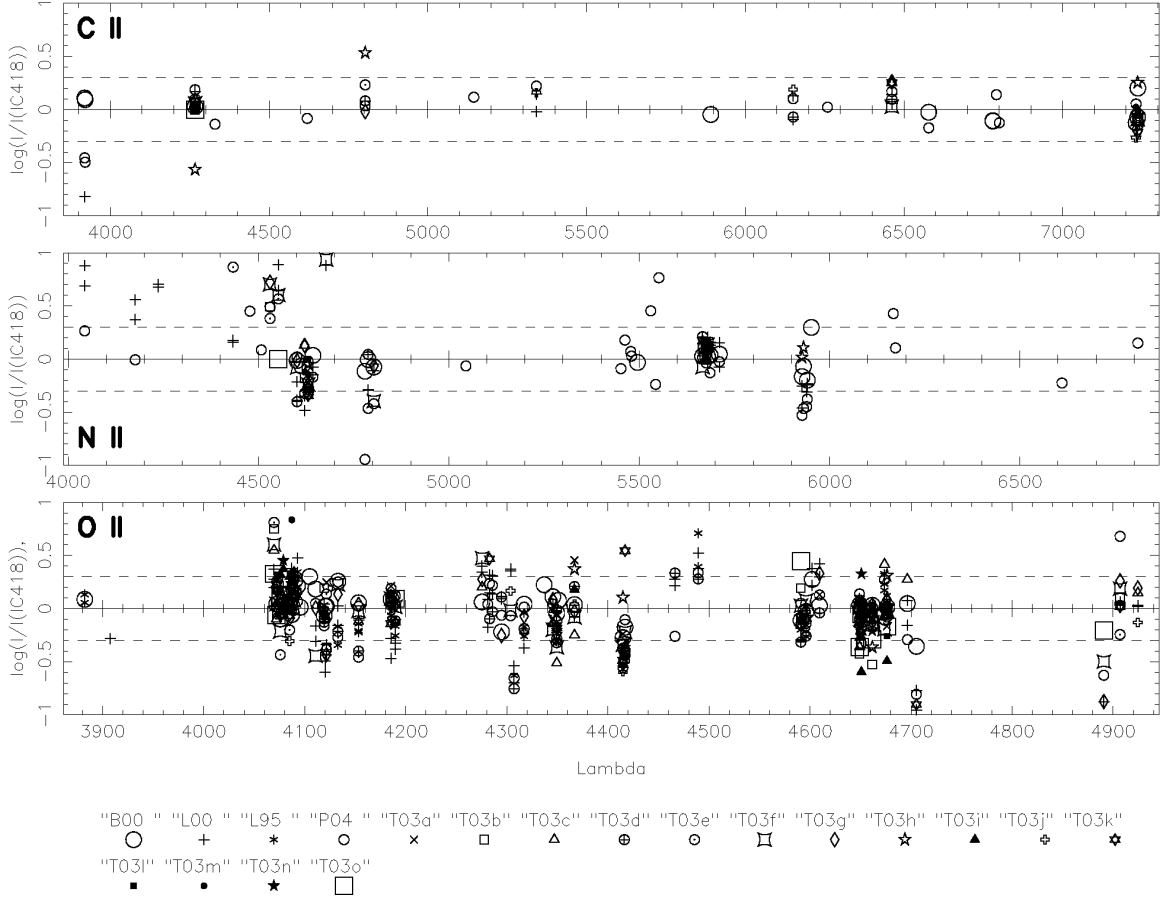}
\caption{Ratios of individual \ion{C}{2}, \ion{N}{2}, and \ion{O}{2}
recombination line intensities from other PNe/H II region spectra with
respect to the same lines observed in IC 418, with normalization and
selection criteria described in the text.  Dashes indicate factor of
two deviations from unity, our conservative limit for observational
errors alone.  The key below the figure matches symbols used in graphs
to their respective objects' literature references: ``BOO'' Orion
Nebula \citep{B00}, ``L00'' NGC 6153 \citep{L00}, ``L95'' NGC 7009
(\ion{O}{2} \textit{only}; Liu et al. 1995), and ``P04'' NGC 5315
\citep{P04}.  In the key those symbols matched to a reference with a
``T03'' prefix refer to objects listed in Tables 3 and 4 of
\citet{T03}: ``TO3a'' NGC 2022, ``T03b'' NGC 2440, ``T03c'' NGC 3132,
``T03d'' NGC 3242, ``T03e'' NGC 3918, ``T03f'' NGC 5315, ``T03g'' NGC
5882, ``T03h'' NGC 6302, ``T03i'' NGC 6818, ``T03j'' IC 4191
(fixed-slit), ``T03k'' IC 4191 (scanning-slit), ``T03l'' IC 4406,
``T03m'' My Cn 18, ``T03n'' SMC N87, and ``T03o'' LMC
N141. \label{f10}}
\end{figure*}

The intensities of these lines are roughly an order of magnitude
less than the stronger recombination lines from the same ions.  Given
the generally weaker strengths of lines which are due primarily to
dielectronic recombination (\ion{C}{2} quartets, \ion{N}{2} quintets,
and \ion{O}{2} sextets specifically), as compared to the other lines
likely to be excited by 2-body recombination alone, it would take a
significant enhancement of dielectronic rates to be competitive with
normal electron recombination for most high level permitted lines.
Dielectronic recombination coefficients from \citet{NS84} are also
available for some multiplets we observed that ordinarily are
considered to be excited purely by one-body radiative recombination:
\ion{C}{2} multiplets V2 and V6, \ion{N}{2} multiplet V3, and
\ion{O}{2} multiplets V1, V2, and V10.  However, at the average IC 418
electron temperature of 10000 K, they all yield values at least an
order of magnitude less than the one-body radiative recombination
coefficients, again necessitating a substantial enhancement in their
rates to influence the intensities of their lines.  Thus, for IC 418
is unlikely that dielectronic recombination is responsible for the
enhanced strength of the high level CNONe permitted lines.

\begin{deluxetable}{lccc}
\tablewidth{3.5in}
\tablecaption{Comparative Ionic Abundances. \label{tab11}}
\tablehead{
\multicolumn{1}{c}{N$^{+i}$/H$^+$} & \colhead{Coll.} & \colhead{Recomb.} &
\multicolumn{1}{c}{Recomb./Coll.} 
}
\startdata
C$^{+2}$/H$^+$ & \nodata   & 5.5(-4) & \nodata \\
N$^+$/H$^+$    & 4.1(-5) & 3.7(-3)\tablenotemark{1} & \nodata \\
N$^{+2}$/H$^+$ & \nodata   & 2.6(-4) & \nodata \\
O$^+$/H$^+$    & 1.7(-4) & 3.4(-4) & 2.0    \\  
O$^{+2}$/H$^+$ & 1.2(-4) & 1.6(-4) & 1.3    \\
Ne$^{+2}$/H$^+$& 4.3(-6) & 2.5(-5)\tablenotemark{2} & 5.8\tablenotemark{2}    
\enddata
\tablenotetext{1}{From continuum fluorescence contaminated lines.}

\tablenotetext{2}{From only two lines with uncertain recombination
coefficients.}

\end{deluxetable}

\section{Ion Abundance Discrepancies in PNe}

A comparison of the ion abundances given in Table~\ref{tab11} for IC
418 reveals that the abundances derived for the ions O$^+$ and
O$^{+2}$ from both forbidden and numerous recombination lines of these
ions agree to within a factor of $\sim$2, and are not as discrepant as
found in other PNe \citep{L95,L00}, where the disagreements typically
exceed a factor of 3.  As mentioned previously, this result for IC 418
is consistent with its high surface brightness.  Various explanations
have been put forward to account for the different abundances derived
from forbidden and permitted lines originating from the same ion.
Lower temperatures enhance radiative recombination, and higher
densities lead to collisional de-excitation of forbidden lines, and
therefore it is possible that either (a) additional excitation
mechanisms such as continuum fluorescence \citep{S68,G75a,G76,E02} and
dielectronic recombination \citep{GD01a} populate the high level
permitted lines, or (b) an inhomogeneous gas having density and
temperature fluctuations \citep{P67,VC94,L00,T04} enhances
recombination line radiation relative to the collisionally excited
lines in the cooler component of the gas.  All of these explanations
for the enhanced strengths of the CNONe permitted lines have been
studied with application to PNe and H II regions, but no consensus has
emerged as to the real cause of the discrepancies.

The recent acquisition of high quality spectra of PNe and H II regions
now provides an excellent database that enables us to test some of the
ideas above.  Normally, recombination lines from the same ion should
have relative intensities with respect to each other that are roughly
the same from object to object, especially lines within the same
multiplet.  This is because recombination coefficients are insensitive
to physical conditions except when densities are sufficiently high to
change the population distribution within the fine structure levels of
the ground state of the parent ion.  This fact constitutes a test for
recombination lines.

We have compared the relative intensities of the high-level permitted
lines of \ion{C}{2}, \ion{N}{2}, and \ion{O}{2} in IC 418 with the
intensities of the same lines observed in the Orion Nebula
\citep{B00}, NGC 7009 \citep{L95}, NGC 6153 \citep{L00}, NGC 5315
\citep{P04}, and selected PNe from the objects studied by
\citet{T03}. We considered only those permitted lines from
Tables~\ref{tab6}--\ref{tab8} which did not have intensity corrections
made to them, to obtain a set of lines with the least ambiguous IDs
and least likely to be affected by blending in the spectrum of IC 418.
This set does, however, include lines which yielded enhanced
abundances relative to other permitted lines in those tables
(i.e. \ion{C}{2} multiplet V3, \ion{N}{2} multiplet V30, etc..).
Additional lines excited by low-temperature dielectronic recombination
from Table~\ref{tab10}, selected under similar criteria as those lines
chosen from Tables~\ref{tab6}--\ref{tab9} were also considered.  For
the lines of each ion in the other objects we normalized the sum of
the line intensities for that ion in the other object to the sum of
the intensities of the same lines in IC 418, and then for each
individual line we determined the ratio $I$(other object)/$I$(IC
418). These ratios should all be close to unity if all the lines are
formed by recombinations. The measured values are presented in
Figure~\ref{f10}, plotted in terms of the wavelengths of the
respective lines.

The dashed lines in Figure~\ref{f10} indicate departures by a factor
of two from a ratio of unity, which we take to be an extremely
conservative limit for the uncertainties of the measured fluxes of
weak lines.  The ratios for \ion{C}{2} are within these uncertainties,
indicating that the \ion{C}{2} relative permitted line strengths are
consistent with recombination, i.e., are stable from object to
object. On the other hand, many of the \ion{N}{2} and \ion{O}{2} lines
have intensity variations that place them outside of the factor of two
range.  The data demonstrate that significant variations in the
relative intensities of supposed recombination lines do exist from
object to object, including large variations of intensities among
lines of the same multiplet.  This suggests that direct recombination
is not the only mechanism populating the high levels of these ions,
although it is possible in a small fraction of objects that densities
could reach levels for which the ground-state fine structure levels
could change from their nominal occupation of the ground level, thus
producing some of the observed intensity variations.  Although the
line profiles of these high level permitted lines are generally
consistent with their parent ion being the next higher stage of
ionization, we conclude that many of these high level permitted lines
are due not to electron recombination but to some other undetermined
process. Note that in some cases the ratios plotted in
Figure~\ref{f10} are significantly less than unity, indicating that
the line in question is relatively stronger in IC 418 than in the
other objects, despite our expectation that the anomalous abundance
effects should be stronger in the other objects.

Dielectronic recombination and charge transfer with H and He are two
of the more likely processes that might augment the population of the
CNONe upper levels, however there are reasons for believing that
neither of these processes are competitive with radiative
recombination.  \citet{S75} and \citet{BD80} have shown that although
charge transfer can populate lower levels at high rates compared to
recombination, for nebular conditions there is too large an energy
threshold for the relevant upper levels of the heavy elements to be
populated by this process.  In addition, the intensities of the
strongest dielectronic recombination lines observed in IC 418 and
listed in Table~\ref{tab10}, which are fed from auto-ionizing levels
that lie just above the ionizing continuum of the parent ion
\citep{NS84} e.g., \ion{C}{2} \lam6780 (multiplet V14), \ion{N}{2}
\lam5535 (multiplet V63), and \ion{O}{2} \lam4465 (multiplet V94), are
all an order of magnitude weaker than the stronger permitted lines
from the same ions.  Thus, it is unlikely that either of these
processes are responsible for the enhanced strengths of the heavy
element permitted lines.

\section{Summary}

From high resolution spectra of the PN IC 418 we have detected and
identified emission lines down to intensity levels less than 10$^{-5}$
that of H$\beta$.  The spectra reveal line profiles that vary greatly
from ion to ion, and that line width is an important parameter to use
in making line identifications and constraining the excitation
processes responsible for a transition.  For this expanding,
photoionized nebula there is a clear relationship between line width
and ionization potential of the emitting ion.  Line profiles take
advantage of the simple kinematics and ionization stratification of IC
418 to isolate the spatial origins of many permitted lines and shed
light onto their excitation mechanisms.  For the temperatures and
densities of the nebular shell we derive ionic abundances for selected
CNONe ions relative to H from the forbidden lines and recombination
lines of these ions.  Consistent with analyses of other PNe we derive
discrepant abundances for the ions from the two types of lines, with
the high level permitted lines indicating abundances somewhat higher
than those derived from the forbidden lines.  The high resolution of
our spectra, the low ionization of IC 418, and our careful
identification of the lines, enable us to rule out line blends as
contributing significantly to the problem of the enhanced permitted
line intensities.

We find that high level permitted lines normally assumed to be excited
by recombination show substantial variations in relative intensity
among different PNe and H II regions, suggesting that other processes
are important in exciting many of these putative recombination lines.
Although dielectronic recombination is a possible cause and has been
suggested as a significant contributor to the permitted lines of CNO
ions, the relatively weak dielectronic recombination lines fed from
auto-ionizing levels argues against this process as a general
explanation for the anomalously strong permitted lines, at least for
IC 418.  The root cause of the anomalous
intensities of the forbidden and permitted lines remains unclear, but
probably involves a number of the suggestions discussed above that
have appeared in the literature.  The one finding that emerges from
this study is the importance for the analysis of emission-line objects
of using a spectral resolution that resolves intrinsic line profiles.

\acknowledgments 

This paper is based upon observations made at the Cerro Tololo
Inter-American Observatory, National Optical Astronomy Observatory,
which is operated by the Association of Universities for Research in
Astronomy Inc., under cooperative agreement with the National Science
Foundation.  BS would like to thank Dr. Peter van Hoof for many
useful discussions.  JAB wishes to acknowledge support from HST grant
GO09736.02-A and NSF grant AST-0305833.

\end{document}